\documentstyle[aps,psfig]{revtex}
\begin{document}
\title{The Image of Self Intersecting QCD Strings in Four Dimensions}
\author{
Bob Bacus$\footnote{bacus@hepaxp.physics.uiowa.edu}$  and V.G.J. Rodgers$\footnote{vincent-rodgers@uiowa.edu}$\\
Department of Physics \\
and Astronomy\\
The University of Iowa\\
Iowa City, IA 52242-1479\\
}

\date{\today}
\maketitle
\begin{abstract}
{\sf
We numerically examine the self-dual solutions of self-intersecting strings 
immersed in four
dimension.  By fixing a frame of reference we are able to show that
self-intersecting {\em open} strings admit world sheet topologies 
that are able to support monopole/anti-monopole
structures. Furthermore we
identify the topologies that can lead to $q - {\bar q}$
production and annihilation and numerically estimate the production rate for
different instantons.  We find an upper bound on the production rate
and use that to estimate the string tension $\mu$.
  By examining the pullback of $\gamma_5$ onto
the string world sheet we show that regions of chiral symmetry
breaking can exist where the string world sheet self-intersects with
itself lending credence to the relationship of self-intersecting
strings and instantons.  We are also able to show that for torus knots, the
intersection number  $\nu=4(p-q)$ which validates an earlier conjecture.
A program written in MAPLE is supplied 
to exhibit the animations of the torus knots in four dimensions. 
These animations are located at the website
http://www-hep.physics.uiowa.edu/\~{}bacus/research.html under the
``Animation Control Panel''.
} 
\end{abstract}
\begin{center}
PACS:  11.25.Pm, 12.38.-t, 12.40.-y \\
Keywords: QCD strings, four dimensional strings, self-intersections
\end{center}

\section{Introduction}

Quantum Chromodynamics (QCD) is widely regarded as the correct theory
of strong interactions.
However, inherently non-perturbative effects such as confinement
remain hidden in the theory due to the nature of the strong coupling
of QCD. In 1973, t'Hooft[\ref{bib:tHooft}] proposed the parameter
$N_c$ of the color gauge group be treated as a free parameter, and 
considered the limit $N_c \rightarrow \infty$ applied to the expansion
of a gauge theory. The resulting expansion gives Feynman diagrams 
which possess the same topology
as the quantized dual string model with quarks at the string ends.
Also, Wilson[{\ref{bib:Wilson}] showed that lattice gauge theory
in the strong-coupling
approximation is a confining theory due to the formation of color charged
strings formed by Faraday flux lines. However this strong coupling approximation
cannot be made in the continuous theory.
Later Makeenko and Migdal[\ref{bib:Makeenko}] as well as 
Gervais and Nevue[{\ref{bib:Gervais}}] were able to establish to
first order in the lattice spacing  an equivalence between a Wilson loop 
\[ \Phi(x^\mu(\tau)) = tr T \exp{(\int A_\nu {d X^\nu \over d\sigma}
d\sigma)} \] and solutions to the single string Schroedinger
equation, provided that the gauge field satisfied the Yang-Mills
equations of motion.  
These indirect probes  suggests that the effective action for QCD
in the strong coupling  regime is a string-like theory.  
Although the precise string theory that governs strongly coupled 
QCD is unknown we expect this string theory to possess certain 
qualitative features that are characteristic of QCD.  
Polyakov[\ref{bib:Polyakov1}] and Balachandran et
al.[\ref{bib:Balachandran}]
examined the effect of adding a term
proportional to the extrinsic curvature to the Nambu-Goto action so
that $\Theta$ vacua effects could be incorporated into the string
ansatz.  The extrinsic curvature term admits self-intersecting 
immersions as solutions and
Nair and Mazur[\ref{bib:Nair}] relate the self-intersections 
number to the instanton number.   

In this note we will examine these self-intersecting
string immersions while in  a fixed
frame in four-dimensions to see if these configurations possess any
realistic topological properties that we expect  from an effective
action of QCD. We will focus our attention  on the torus knot solutions of
Robertson[\ref{bib:Robertson}] in order to be explicit.  
We will show that there exists topologies that 
supports monopole/anti-monopole fields whose flux is constrained to a
finite string (tube) and in some cases to strings that are infinite in
length.  Furthermore we will show that there exists topologies that
can support pair production and annihilation.  By this we mean that
the immersed world sheets have remnants that appear as finite segments
of strings that emerge and dissolve in a fixed frame. These segments are
related to an integer $q$ that appears in the self-dual solutions that
we consider. 
 We are able to show that $q=111$ puts an upper limit on the production
 of fermions.  We then
examine the pullback of $\gamma_5$ onto the world sheet and show that
there can be chiral symmetry breaking ``bubbles'' around the points
where the immersion self-intersects with itself.  This helps
strengthen the relationship between intersecting strings on the one
hand and QCD instantons on the other.  Next we verify the conjecture
of Robertson [\ref{bib:Robertson}] by showing that for torus knot
solutions that the intersection number is given by $\nu = 4(p-q)$. 
We end by providing a program in {\bf MAPLE } that annimates the 
torus knot and torus link solutions.

\section{Topological Support For QCD Processes From  Self-Dual Strings }

One of the questions we ask is what topologies are provided by the
string to support the gauge fields and quarks in four dimensions. 
We are interested in the classical features of the self-intersecting
instantons since these feature should appear as first order quantum
corrections to the effective action of QCD in the strong coupling
regime.  In order to appreciate
the behavior and support of the strings we need to sit in a frame and
examine the topologies that the immersed string unfolds.   Since the
self-dual solutions have non-trivial winding number any features we
find related to the self-intersection will be protected from quantum
fluctuations since winding number is a topological propoerty of the
string configuration.   To proceed we will look at
the Nambu-Goto action and include modifications introduced  by Polyakov.  
In order to minimize the contributions of crumpled string world sheets
to the effective action of QCD, Polyakov proposed the action following;
\begin{equation}
S = \mu \int \sqrt{g} d\,\sigma d\,\tau + \frac{1}{\alpha_0} \int \sqrt{g} K^{Aa}_b K^b_{Aa} d \sigma d \tau
\end{equation}
where $K^{Aa}_b$ is the extrinsic 
curvature tensor. However the presence of the extrinsic curvature term not
only  acts as 
a regulator [\ref{bib:Durhuus},\ref{bib:Polyakov1}] for bosonic strings but also provides
an avenue for the introductions of the topological instanton
solutions.
Furthermore it has been shown that the presence of the extrinsic
curvature is necessary for the effective action of free fermions
projected onto the world sheet[\ref{bib:Parthasarathy}].  

The last summand to the above action 
can be expressed in an alternate form as;
\begin{eqnarray}
S_1 &=& \frac{1}{\alpha_0} \int \sqrt{g} g^{ab} \partial_a t_{\mu \nu}
\partial_b t^{\mu \nu} d \sigma d \tau \\
S_1 &=& \frac{1}{2 \alpha_0} \int \sqrt{g} g^{ab} ((\partial_a t_{\mu \nu} \mp
\partial_a \tilde{t}^{\mu \nu})(\partial_b t_{\mu \nu} \mp
\partial_b \tilde{t}^{\mu \nu}) \pm (\partial_a \tilde{t}_{\mu \nu}
\partial_b t^{\mu \nu})) d \sigma d \tau \\
& &\mbox{where } t^{\mu \nu} = \frac{\epsilon^{ab}}{\sqrt{g}} \partial_a X^{\mu}
\partial_b X^{\nu} \label{eq:dual}
\end{eqnarray}
As with QCD, the action is minimized by searching for self-dual solutions, viz.
$\partial_a t^{\mu \nu}=\partial_a {t^*}^{\mu \nu} $. Wheater[\ref{bib:Wheater}]
showed that any surface embedded in four-dimensional space-time which is a
complex curve is a solution to the equations of motion for the extrinsic
curvature term. Robertson[\ref{bib:Robertson}] showed one explicit example of
a self-dual string instanton is a (p,q) torus knot. These are examples
of superminimal immersions into $R^4$[\ref{bib:Landolfi}]. Pawelczyk
has extended the classification of instantons and includes the Euler
character of the surface as well [\ref{bib:Pawelczyk}].  A torus is
parametrized by two variables, $u$ and $v$, representing local coordinates
on the surface. The torus knot is the curve on the torus defined by
the constraint $u^p + v^q = 0$. Figure 1 shows an example, the (3,2) torus
knot.
\begin{figure}[h]
\vspace{1.in}
\centerline{\psfig{file=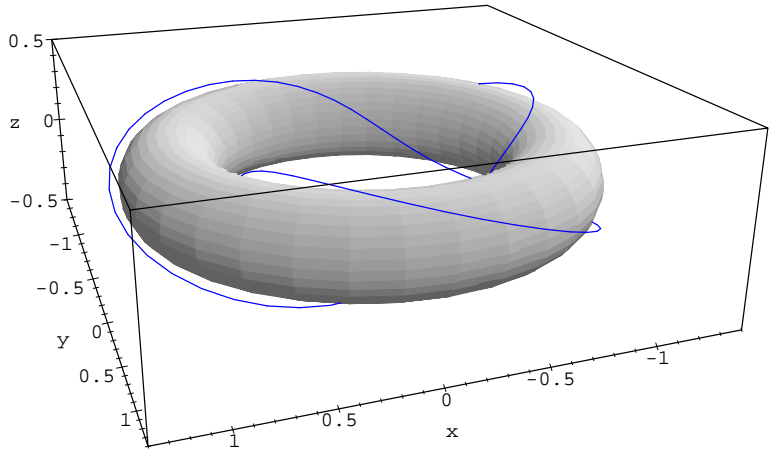,width=4.0in}}
\caption[The (3,2) torus knot]{The (3,2) torus knot.}
\end{figure}

For a (p,q) torus knot instanton solution the string vector is
\begin{equation}
X^{\mu} = [\Im (z^p), \Re (z^p), \Re (-z^q), \Im (-z^q)], 
\end{equation}
where $z = \tau + i \sigma$ and $\Im $ and $\Re$ are the real and
imaginary parts respectively.
Note that Robertson has shown somewhat more generally that a string vector
of the form
\begin{eqnarray*}
X^{\mu} = [\Im (F(z)), \Re (F(z)), \Re (G(z)), \Im (G(z))] \\
\end{eqnarray*}
with $F(z)$ and $G(z)$ as any functions analytic in $z$ are also solutions
to the equations of motion.
These solutions automatically satisfy the Euclidean string equations 
\begin{equation}
 \partial_\tau \partial_\tau X^\mu + \partial_\sigma \partial_\sigma
X^\mu =0, 
\end{equation}
as well as the Euclidean constraint equations 
\begin{equation}
 \partial_\sigma X^\mu
\partial_\tau X^\mu = 0,
\end{equation} and 
\begin{equation}
(\partial_\sigma X^\mu)^2 = 
(\partial_\tau X^\nu)^2.
\end{equation}  

To study these solutions, we create computer animated plots of the solutions
over a finite time interval. We start with a particular (p,q) torus knot
solution such as the (3,2) knot.   This is a topologically
non-trivial solution with self-intersection number $\nu = 4$. The solution
takes the form
\begin{eqnarray*}
X^{\mu}(\sigma, \tau) = [3 \tau^2 \sigma - \sigma^3, \tau^3 - 3 \tau \sigma^2,
 \sigma^2 - \tau^2, -2 \tau \sigma] = [x, y, z, t]
\end{eqnarray*}
$X^{\mu}$ is a mapping from ($\sigma,\tau$) space to ($x,y,z,t$) space.
We choose the last coordinate to be our time parameter and invert the 
the last term and replace $\tau$ in $X^{\mu}(\sigma, \tau)$ with the solution
in terms of $X^4=t$ and $\sigma$. (One may chose other coordinates for
the time such as $X^2$ where the $p=1$ cases correspond to the usual
choice where $\tau$ is the time parameter.) 
 In the case of the (3,2) torus knot, the
inversion of the last term gives $\tau = - \frac{t}{2 \sigma}$; substitution
into $X^{\mu}$ gives
\begin{eqnarray*}
X^{\mu}(\sigma, t) = [\frac{3 t^2}{4 \sigma} - \sigma^3, -\frac{t^3}
{8 \sigma^3} + \frac{3 t \sigma}{2}, \sigma^2 - \frac{t^2}{4 \sigma^2}, t].
\end{eqnarray*}
For each value of the time $X_4=t$ we generate a plot of the string. These
plots are combined to form the animation. The ranges chosen
were $\sigma = [-1,1]$, $t = [-1, 1]$.  The open string is
symmetric in our case but we are careful not to include
characteristics that are due to this symmetry of the solution. We have
marked the image of $\sigma = 1$ and $\sigma = -1$ in order to keep
track of the string endpoints on the Reimann sheet..

There are several characterisitics that we observed:
\begin{itemize}
\item As opposed to finite string segments, the majority of sigments
  are semi-inifinte in length.  This is due mainly to the inversion of
  the time parameter.  We could have chosen $X^2$ to be
  the time component.  Then for the small sector corresponding to
  $p=1$, $t$ would be $\tau$.  However all other sectors would exhibit
  the presence of semi-infinite segments.  These segments may be
  thought of as flux tubes that carry the monopole and anti-monopole 
  gauge fields from a point   out to infinity.  One can construct 
  gauge invariant objects that can
  live on these topologies 
  (see Eq.[\ref{eq:wilsonline1},\ref{eq:wilsonline2}]).
\item The strings interact by exchanging flux lines.  Whenever two
  semi-infinite segments, say A and B,  touch the result is that A
  will pass its segment above the intersection point onto B while
  sewing the segment that B had above the intersection point onto its
  lower portion.   This exchange of flux lines can also take place
  with segments that are finite.  
\item For the case when $p$ is even and $q=2$ 
 then the image is that of a single
  semi-infinite segment that is actually the image of the string
  projected back onto itself.  In this case if one had assigned a
  quark at $\sigma = 1$ and an anti-quarks at $\sigma = -1$ then the
  endpoint of this type of image would correspond to a ${\bar q}q$
  bound state with gauge field lines moving up and moving down the
  flux line that goes out to spatial infinity effectively canceling
  the contribution of the gauge field in the Wilson line (see
  Eq.[\ref{eq:wilsonline1},\ref{eq:wilsonline2}]).   Such configurations
  are expected to contribute to the $<{\bar q} q >$ condensate.
  Examples of this are the (2,2), (4,2), (6,2) and (8,2). 
\item One of the most interesting observations is the appearance and
  disappearance of finite segments of lines.  These segments can
  appear at one point, interact with some segment by exchanging flux
  lines and then disappear at another point.  The emerging segment
  can  have a ${\bar q} q$ pair attached to the ends providing
  topology for pair productions and pair annihilation.  We will
  discuss this process in some detail shortly. Nice examples are (4,3),
  (5,3) and (1,3) torus knots.  
\item  Surfaces that do not intersect $p=q$, appear as infinite or semi-infinte
  segments.  The multiple images of the endpoints can be seen embedded
  in the string segment itself undergoing pair production and
  annihilation along the flux tube itself.  This activity can also be
  seen in torus knots were $p$ is not relatively primed to $q$.  
 Examples are (6,3), (4,2), and (8,4).  
\item Finally we note that the image of $\sigma = \pm 1$ does not
  always map into any endpoint of the image.  
\end{itemize}
 
\begin{figure}[h]
\vspace{.5in}
\centerline{\psfig{file=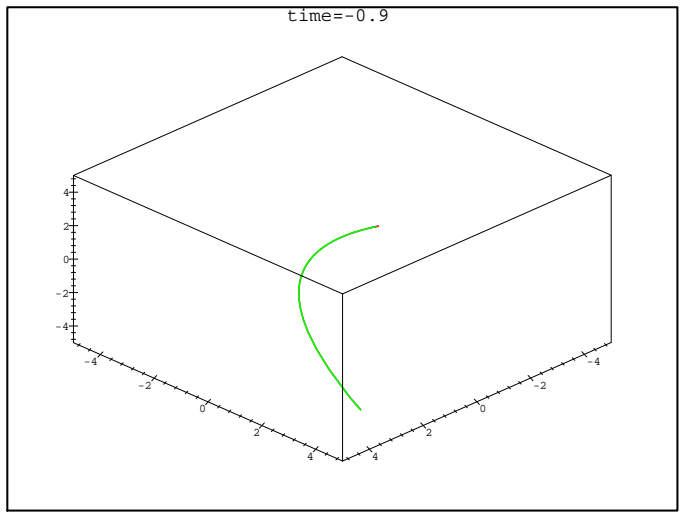,angle=-90,height=3.0in}}
\caption[The (4,2) torus link instanton at time index t=-0.9]
 {The (4,2) torus link instanton at time index t=-.9. 
\\When $p(>q)$ and $q$ and are relatively unprimed, a single semi-infinite
segment appears. This is actually the image of two semi-infinite strings that
mapped on top of each other.}
\end{figure}
  
\eject

\begin{figure}[h]
\vspace{1.in}
\centerline{\psfig{file=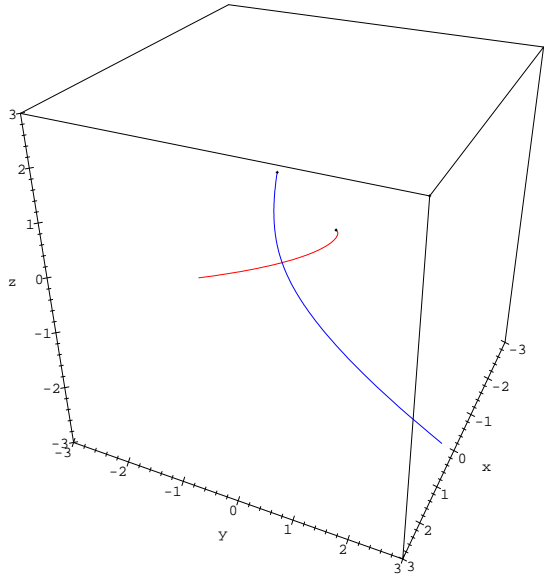,height=3.0in}}
\caption[The (3,2) torus knot instanton at time index t=0.60]
 {The (3,2) torus knot instanton at time index t=0.60}
\end{figure}
Figure 3 shows the (3,2) torus knot instanton at time $t = 0.60$. The
ends of the string are labeled with points, and the string is colored
blue to represent $\sigma > 0$ and red to represent $\sigma < 0$ in order
to visually distinguish these parts. The bounding box containing the plot
has dimensions $x = -3..3, y=-3..3, z=-3..3$. In this frame we see the string
broken in two halves; where  part of the
string around $\sigma = 0$ leaves the bounding box and is mapped out
to infinity.
 \begin{figure}[h]
\vspace{1.in}
\centerline{\psfig{file=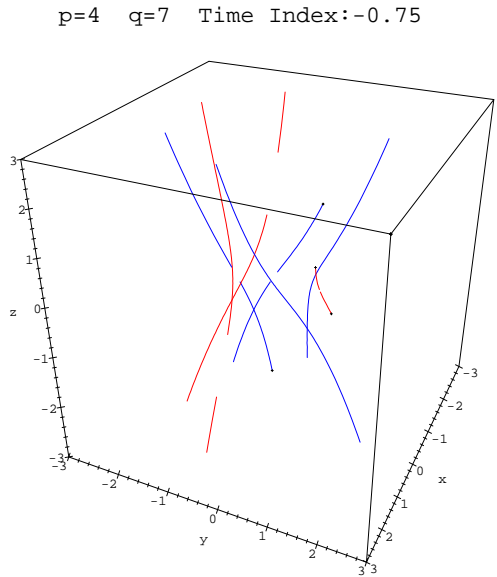,height=3.0in}}
\caption[The (4,7) torus knot  instanton at time index t=-0.75]
 {The (4,7) torus knot instanton at time index t=0.60}
\end{figure}
Here is an example of a knot with negative intersection number.  
The vertical strands are 
semi-infinite in length. Here the broken strands are an imaging artifact.
Just right of the center one can see the
emergence of a finite segment. This segment will go on to exchange
flux lines with the semi-infinite flux lines and then collapse back
into the vacuum. These finite strands  provide support for 
pair production and annihilation.
\eject
\begin{figure}[b]
\begin{picture}(100,200)(60,600)
\put(50,600){\psfig{figure=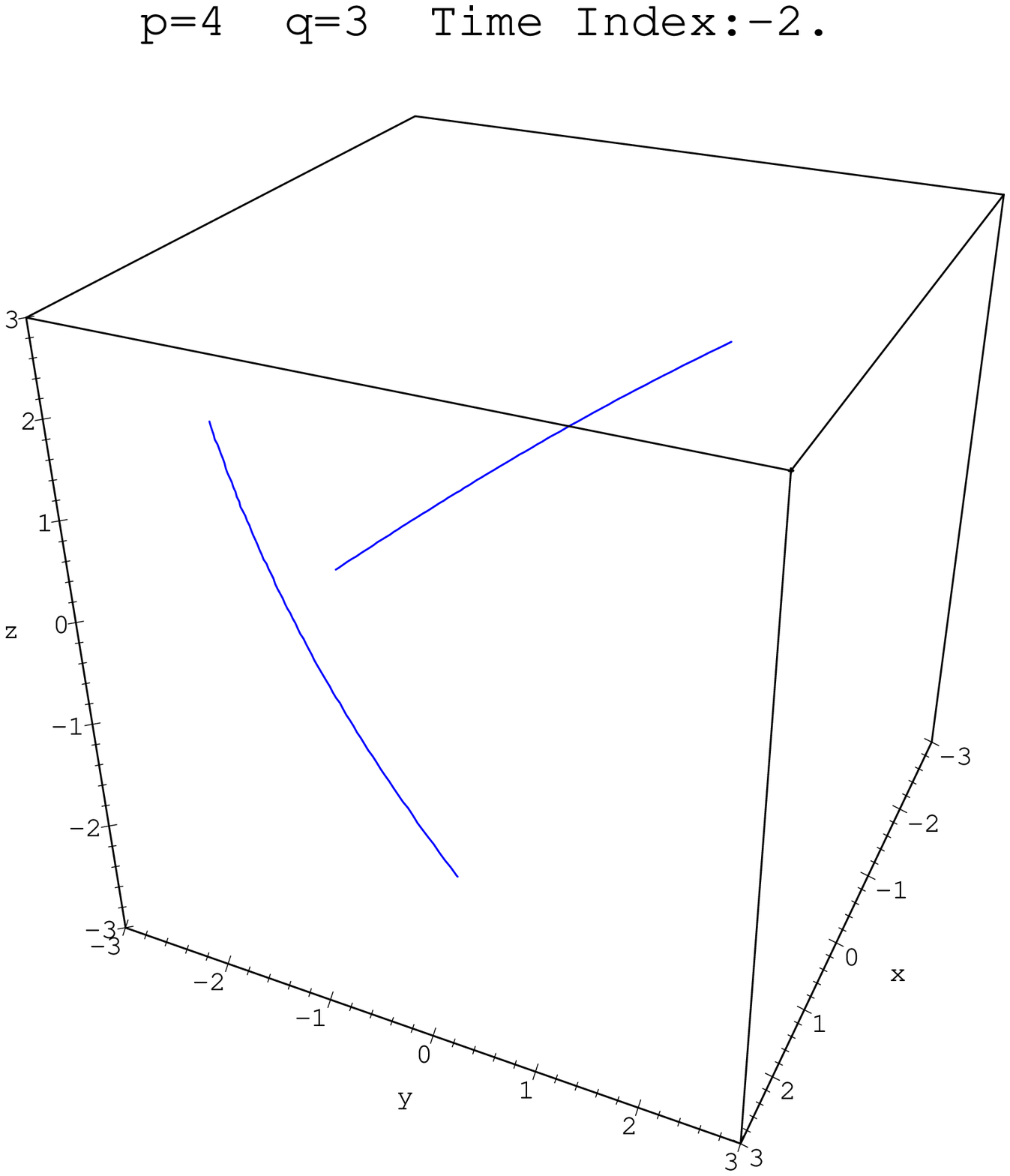,height=2.0in,width=2.0in}}
\put(220,600){\psfig{figure=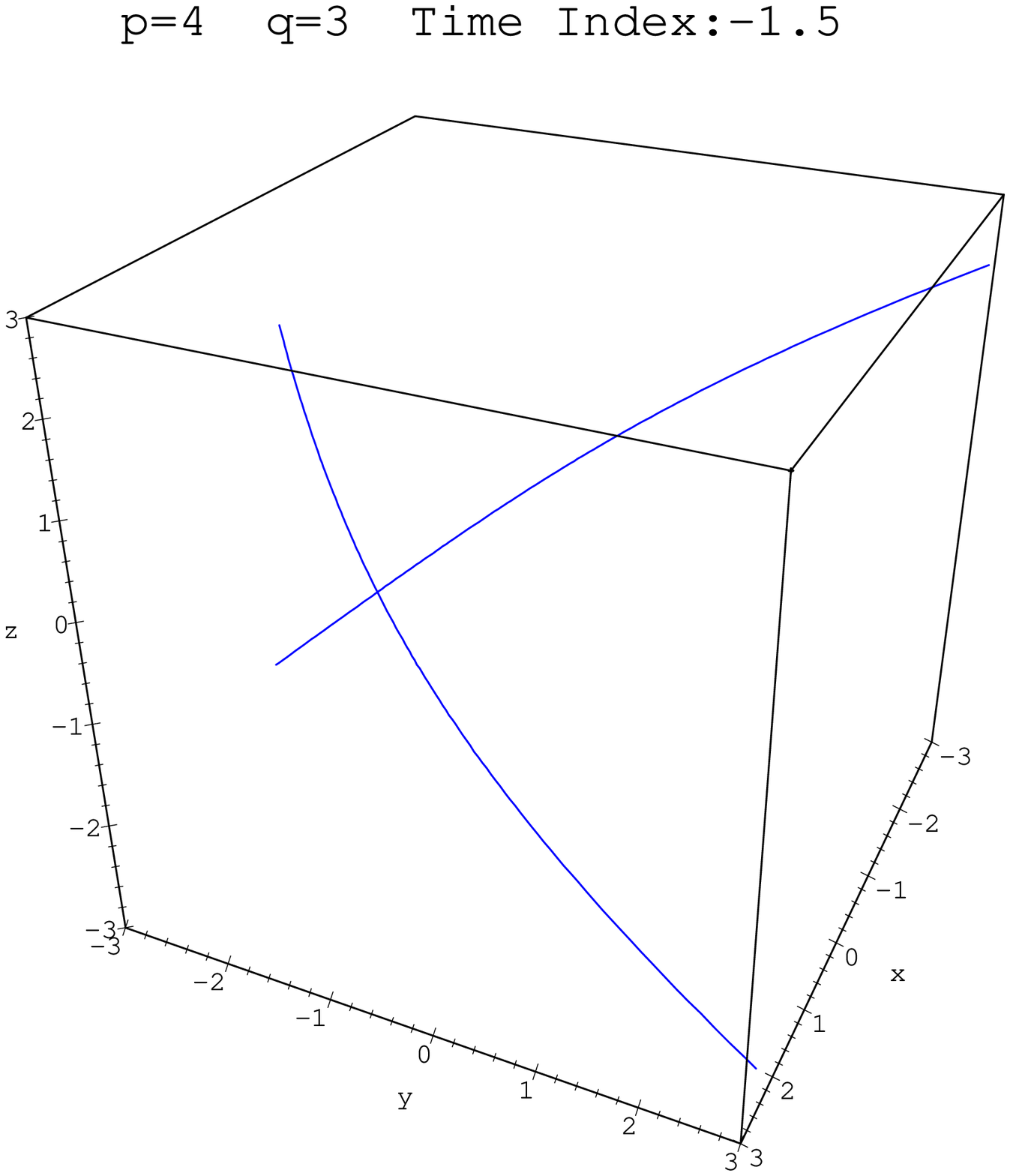,height=2.0in,width=2.0in}}
\put(390,600){\psfig{figure=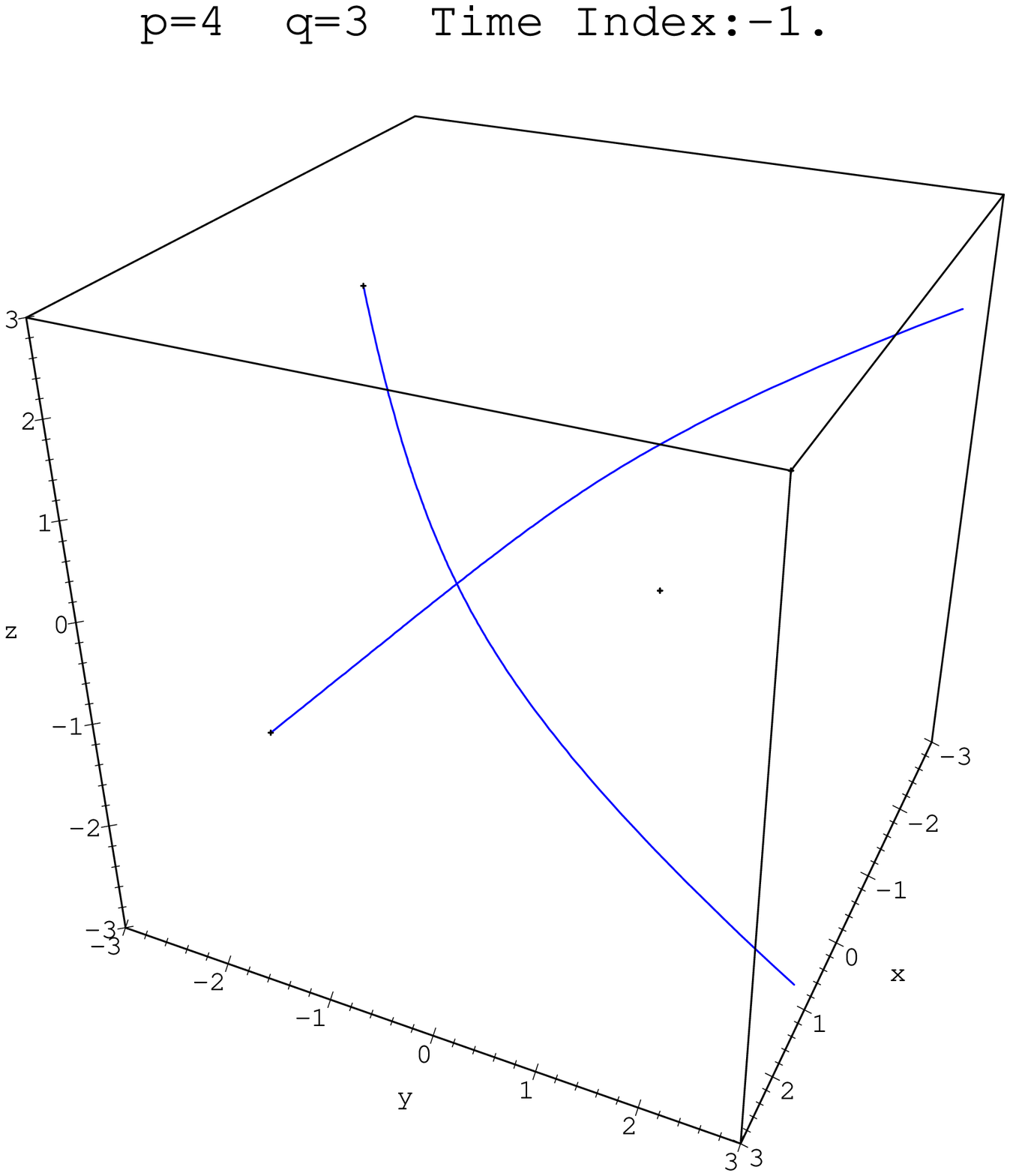,height=2.0in,width=2.0in}}
\put(50,400){\psfig{figure=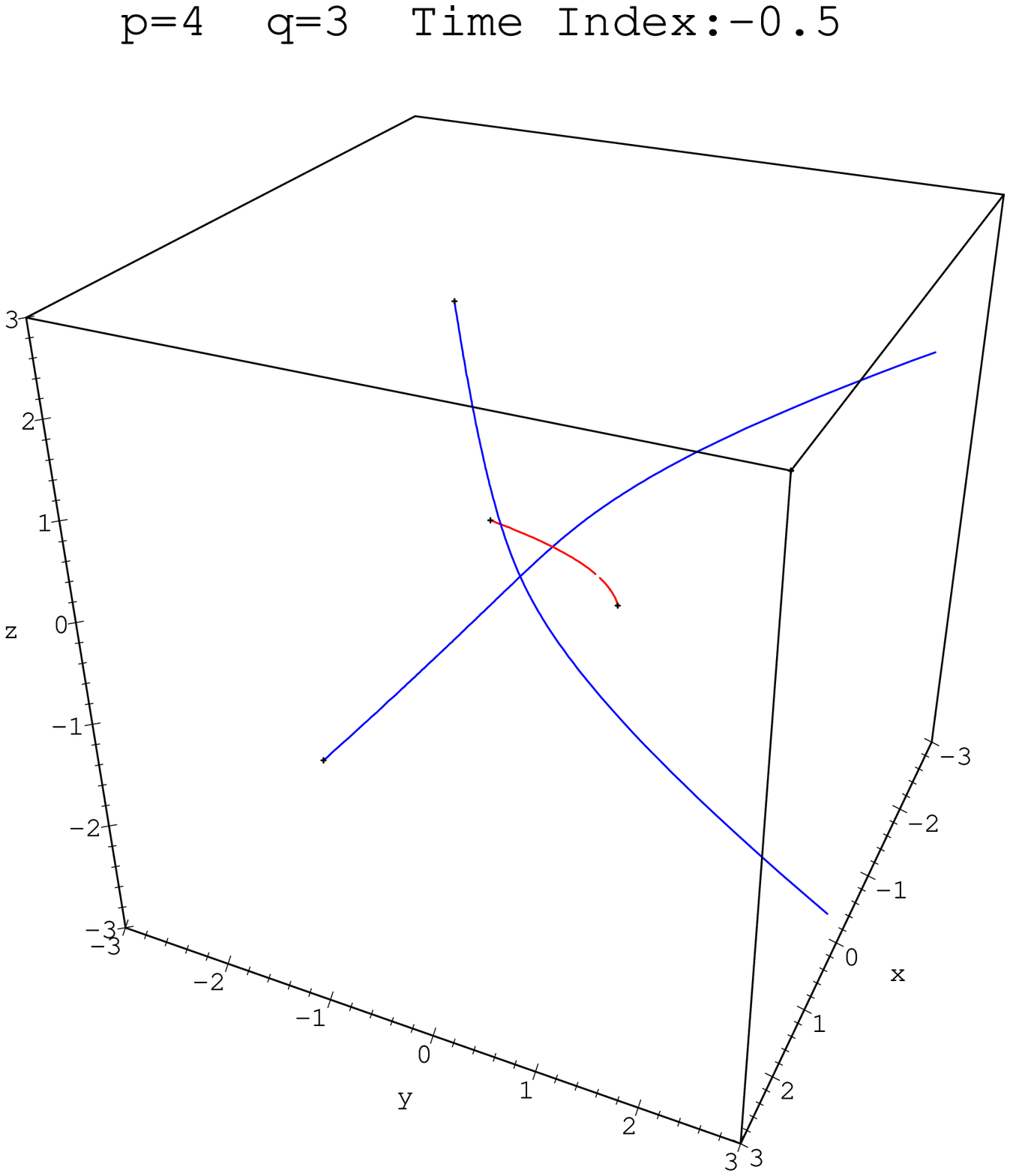,height=2.0in,width=2.0in}}
\put(220,400){\psfig{figure=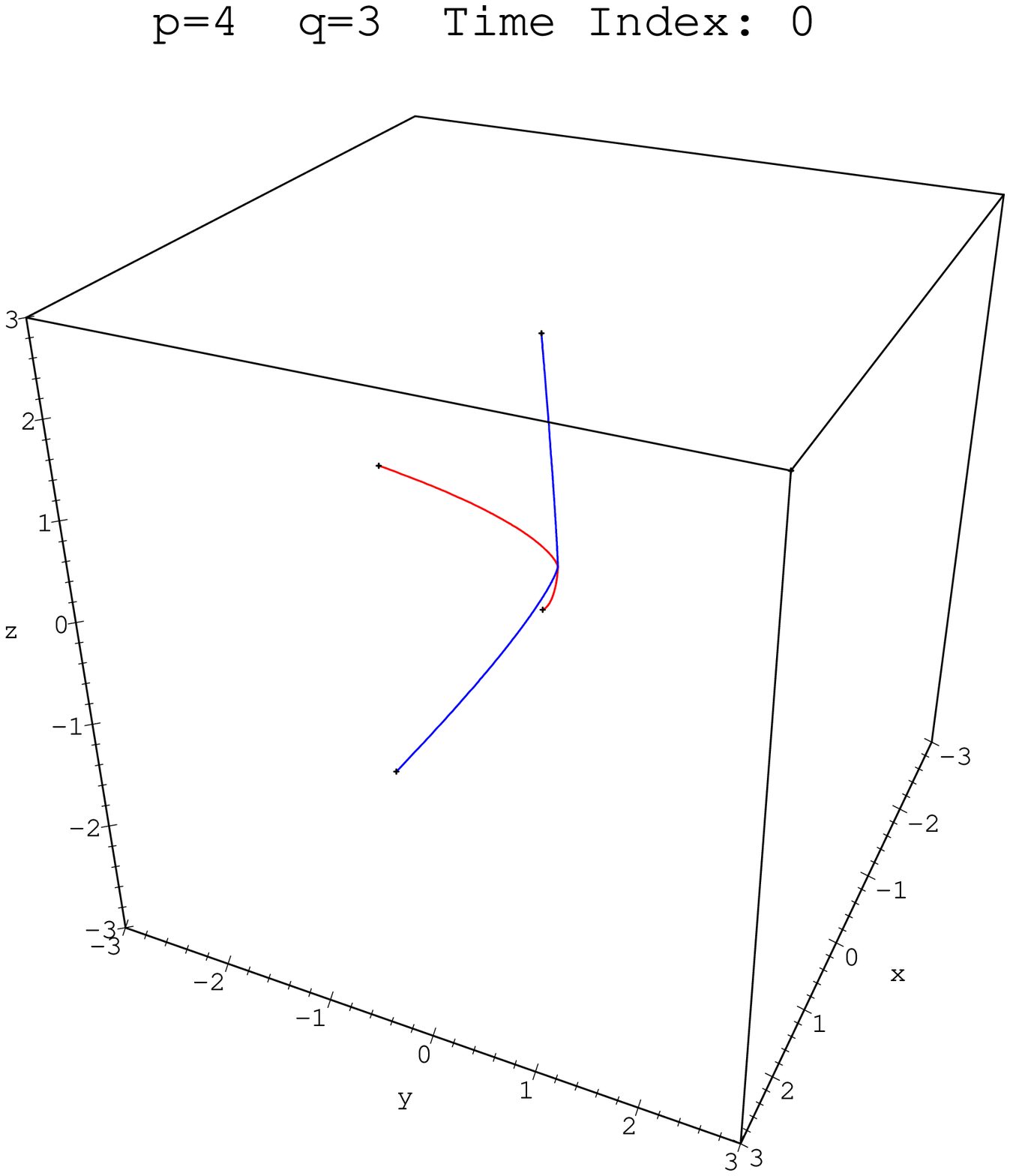,height=2.0in,width=2.0in}}
\put(390,400){\psfig{figure=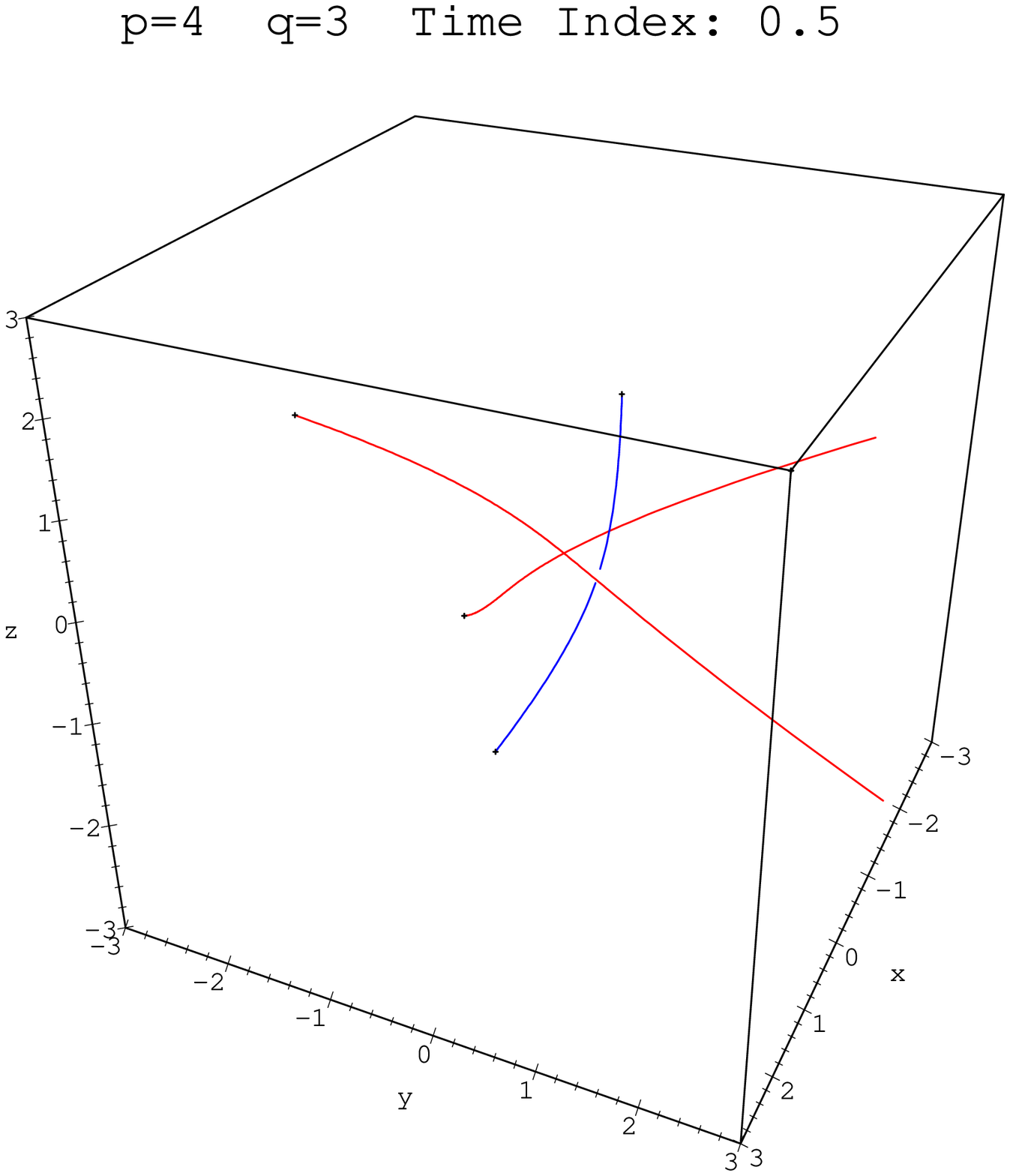,height=2.0in,width=2.0in}}
\put(50,200){\psfig{figure=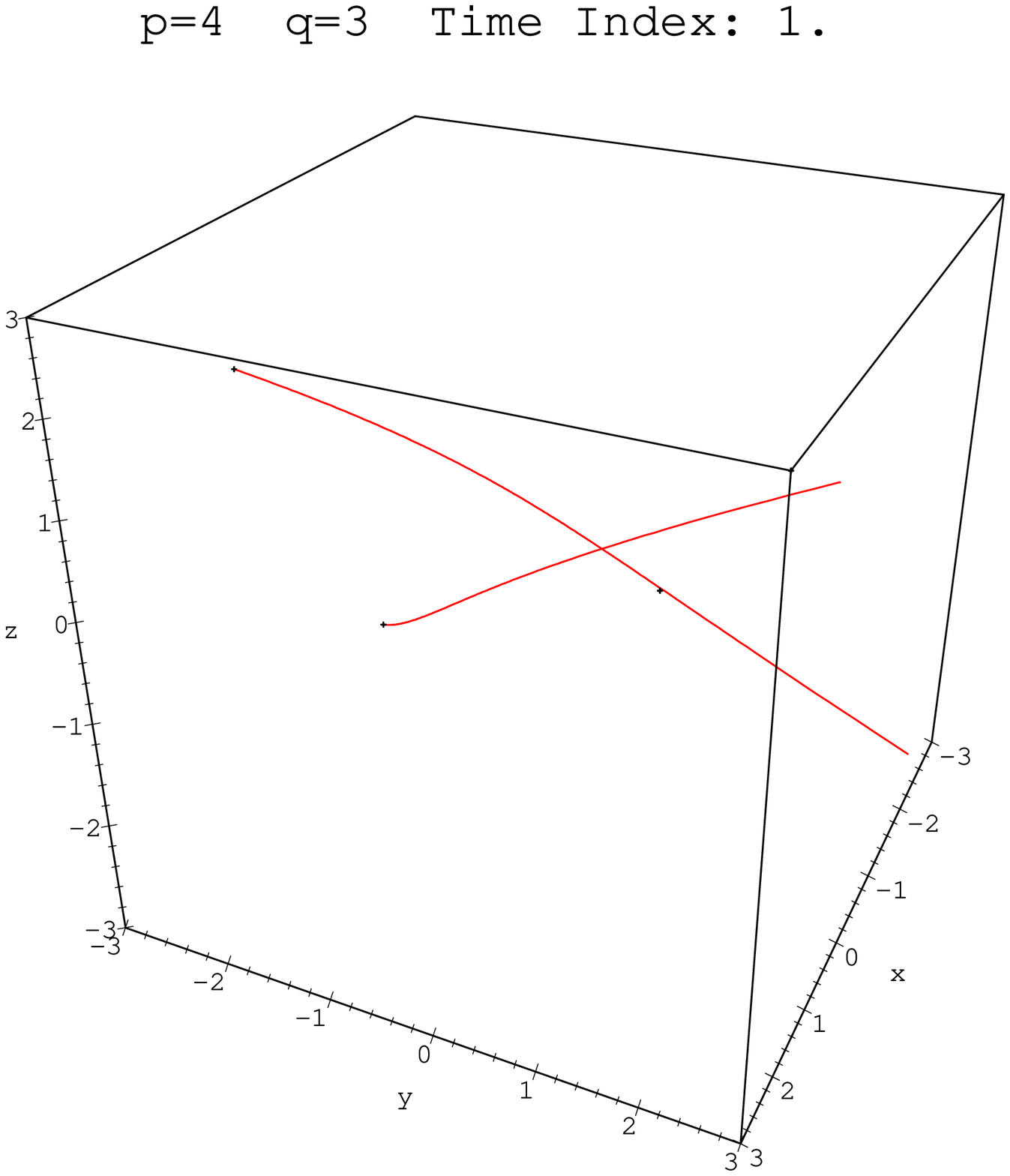,height=2.0in,width=2.0in}}
\put(220,200){\psfig{figure=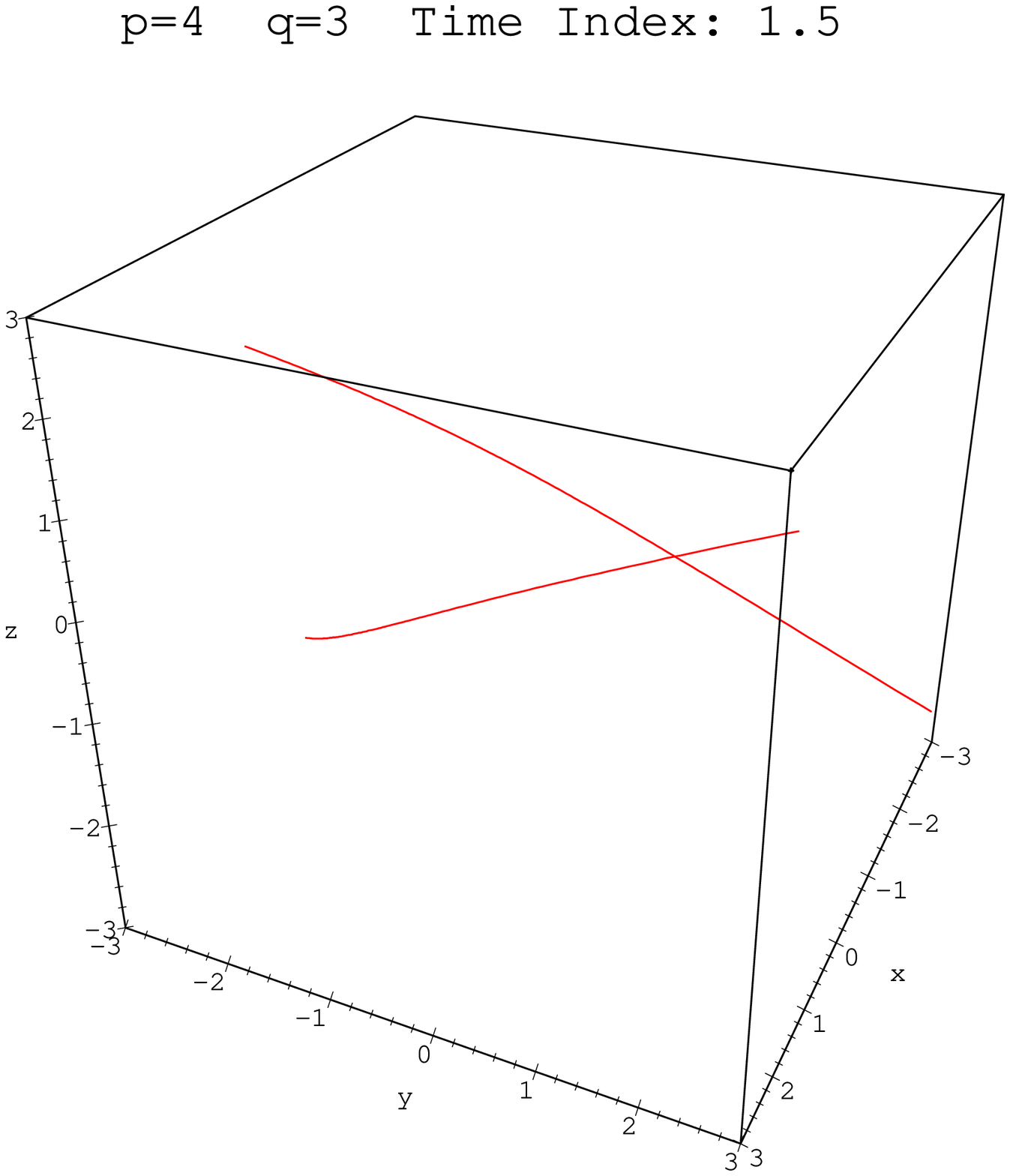,height=2.0in,width=2.0in}}
\put(390,200){\psfig{figure=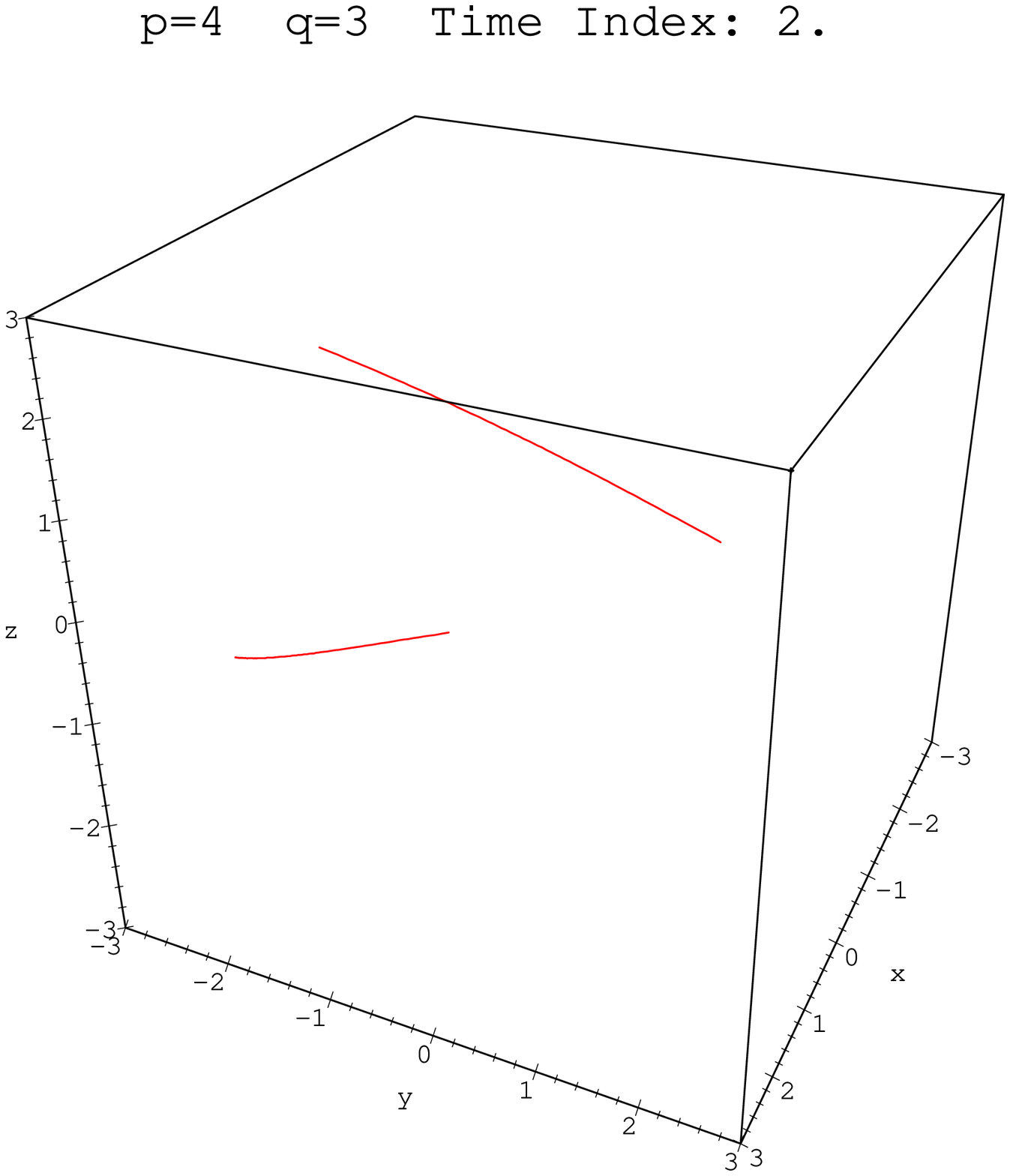,height=2.0in,width=2.0in}}
\end{picture}
\vskip6.0in
\caption[Evolution of the (4,3) torus knot instanton.]
{Evolution of the (4,3) torus knot instanton.}
\end{figure}
\eject
\begin{figure}[b]
\begin{picture}(100,200)(60,600)
\put(50,600){\psfig{figure=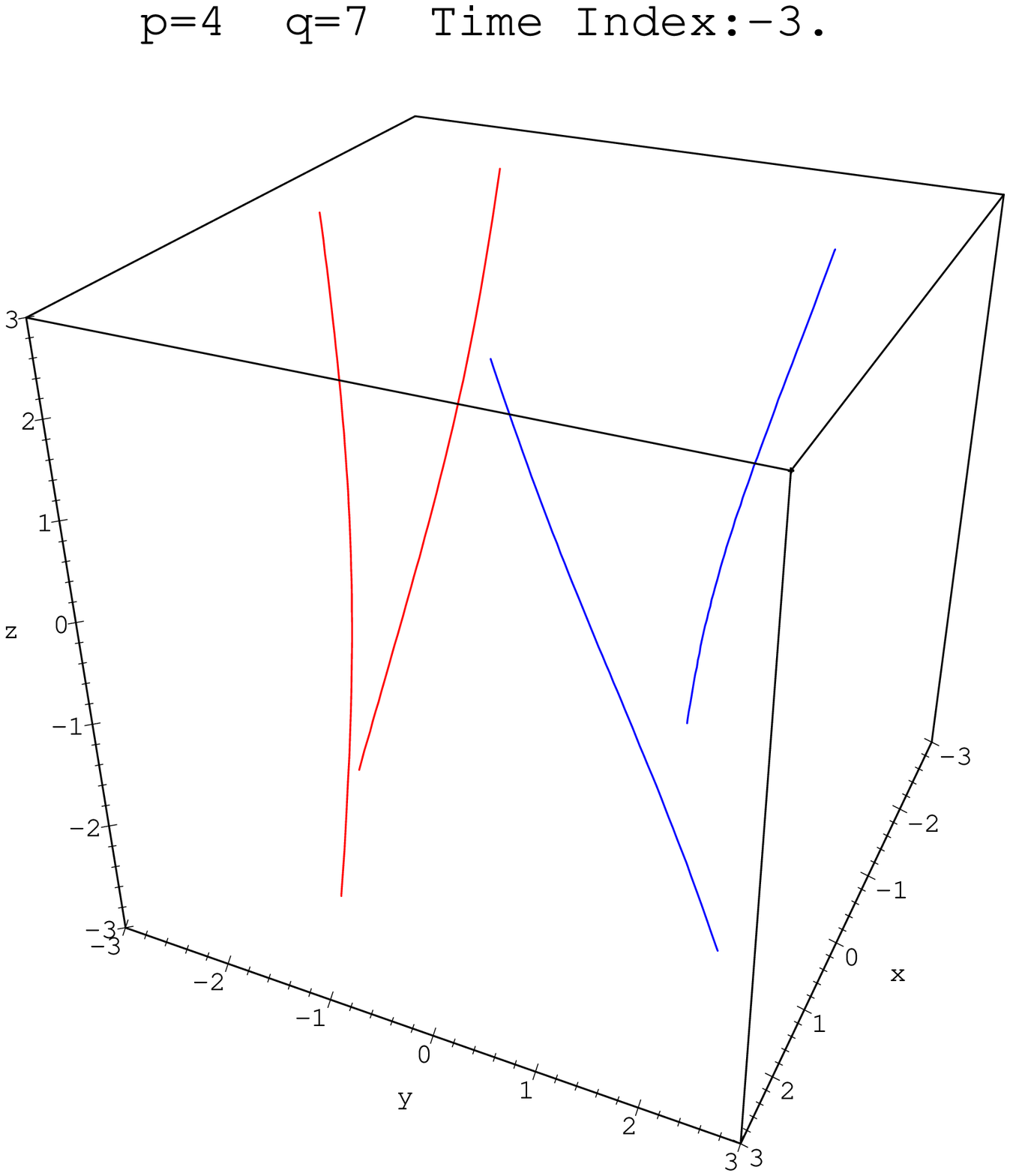,height=2.0in,width=2.0in}}
\put(220,600){\psfig{figure=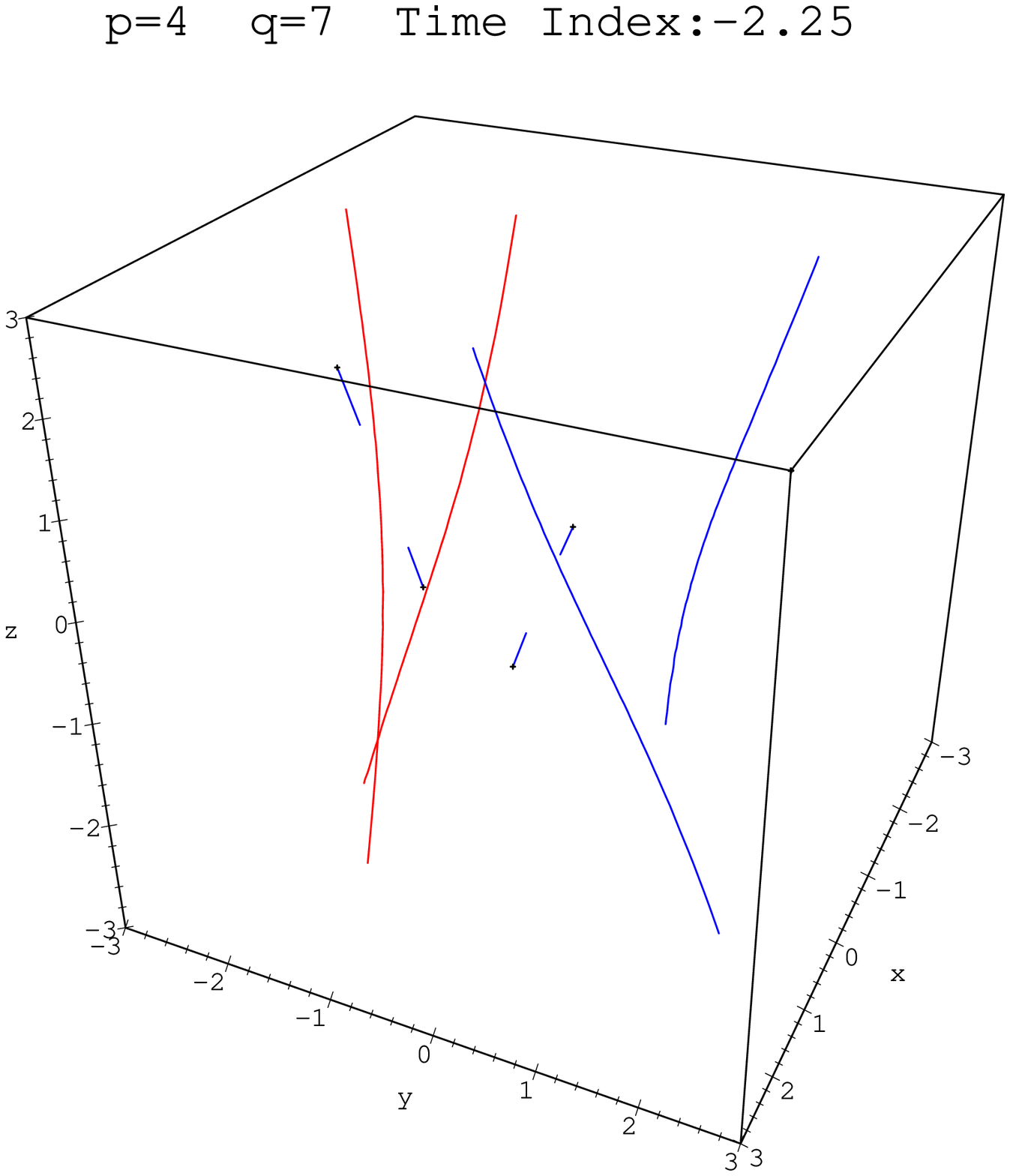,height=2.0in,width=2.0in}}
\put(390,600){\psfig{figure=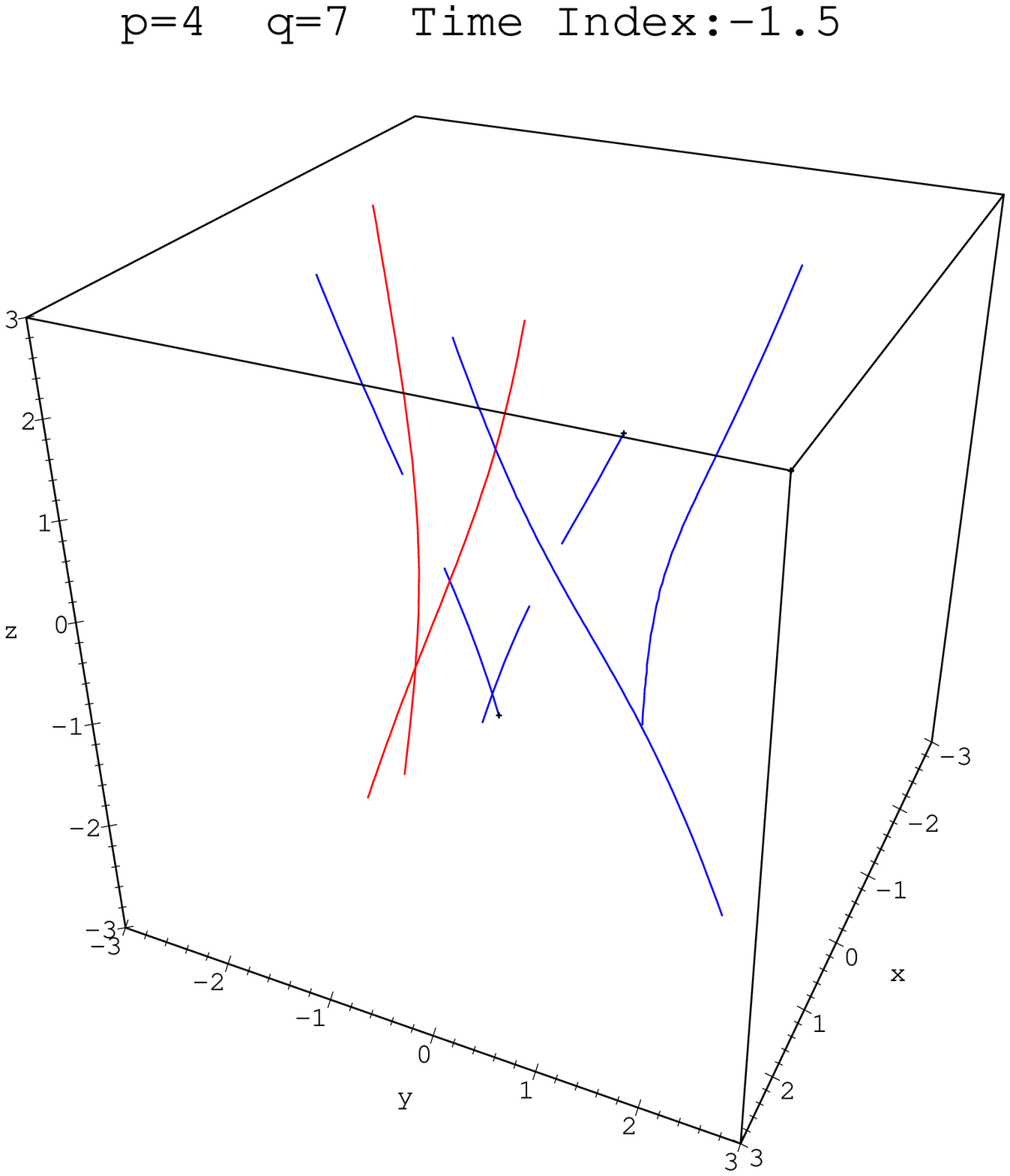,height=2.0in,width=2.0in}}
\put(50,400){\psfig{figure=4x7-3.eps,height=2.0in,width=2.0in}}
\put(220,400){\psfig{figure=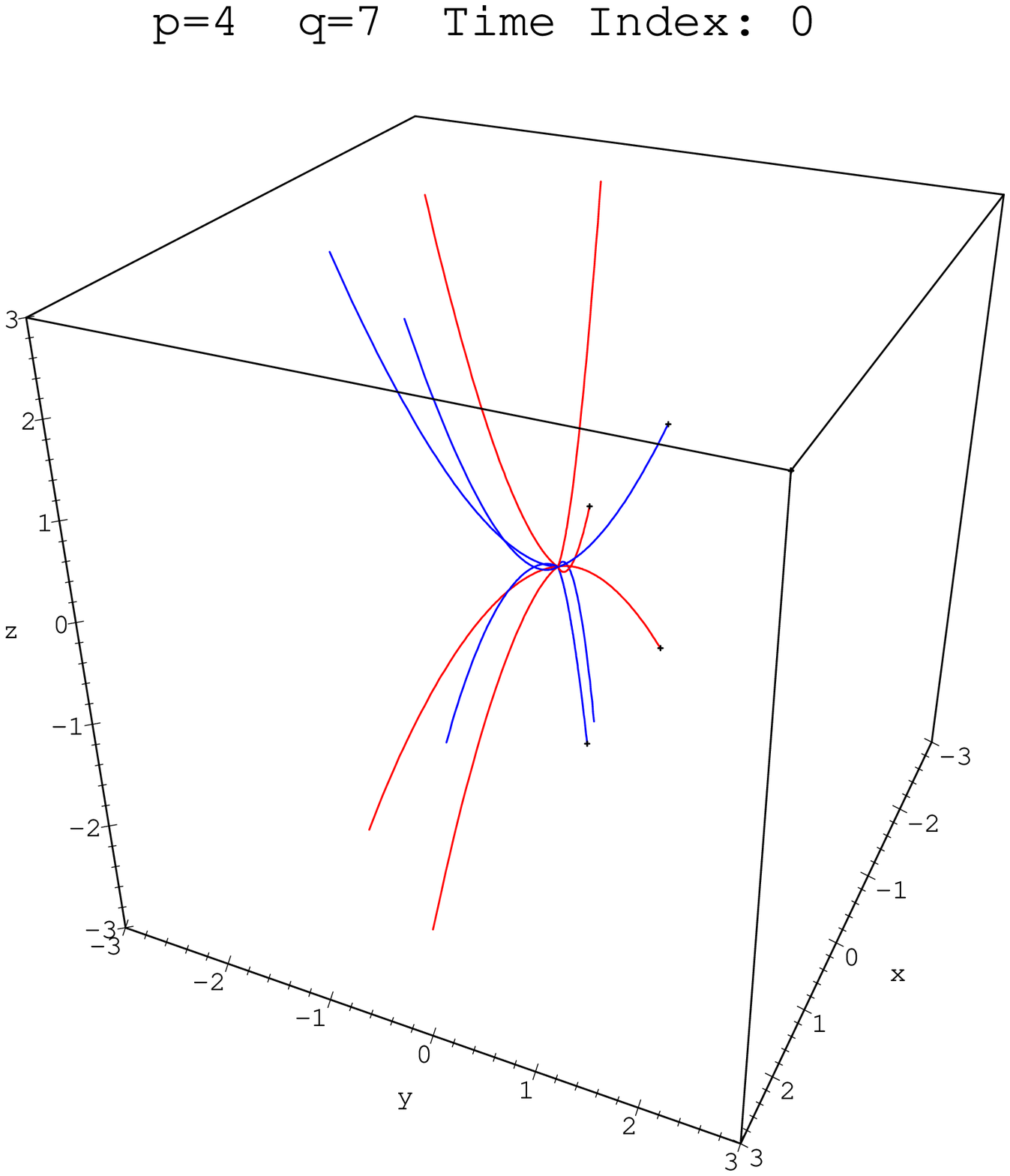,height=2.0in,width=2.0in}}
\put(390,400){\psfig{figure=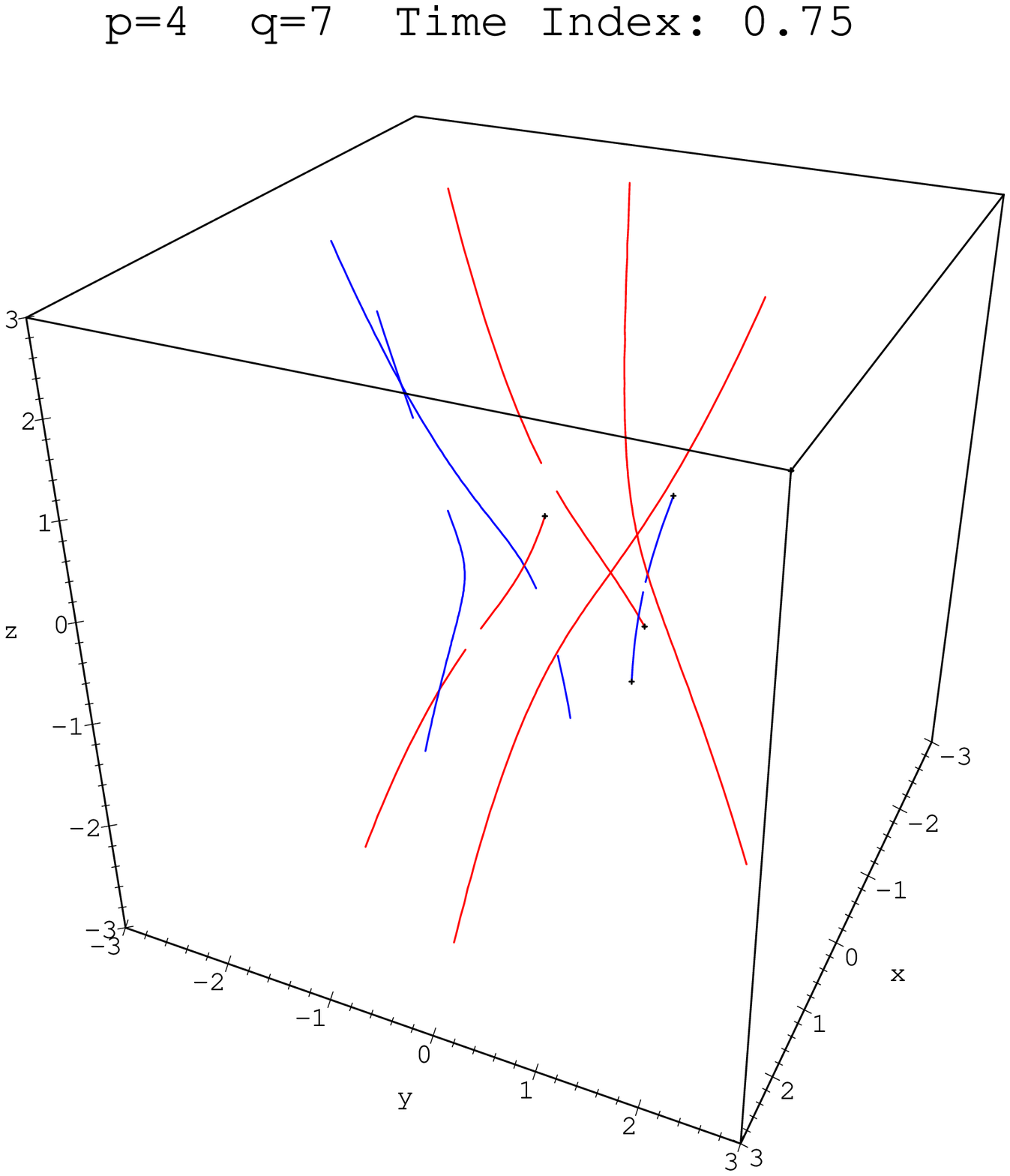,height=2.0in,width=2.0in}}
\put(50,200){\psfig{figure=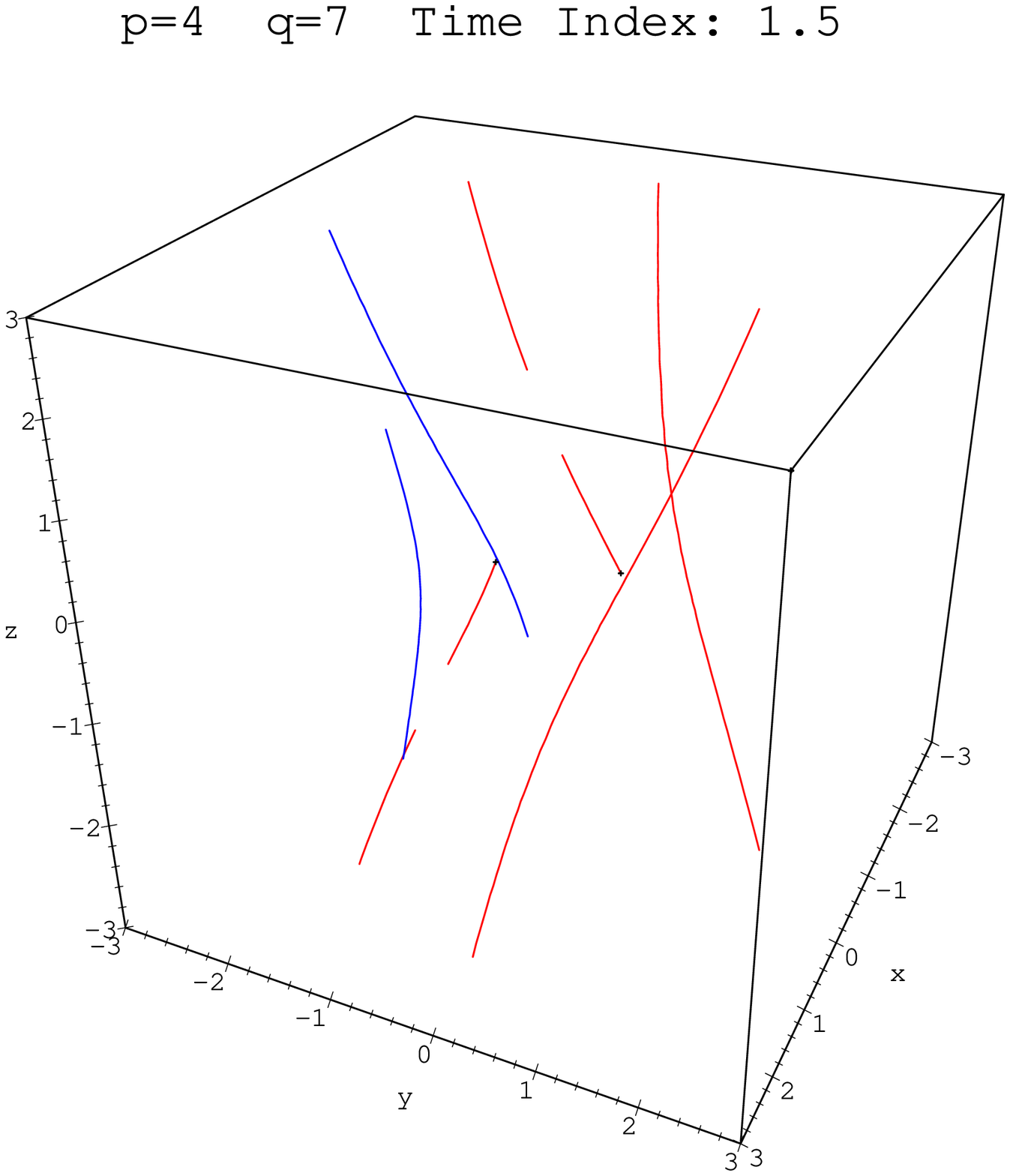,height=2.0in,width=2.0in}}
\put(220,200){\psfig{figure=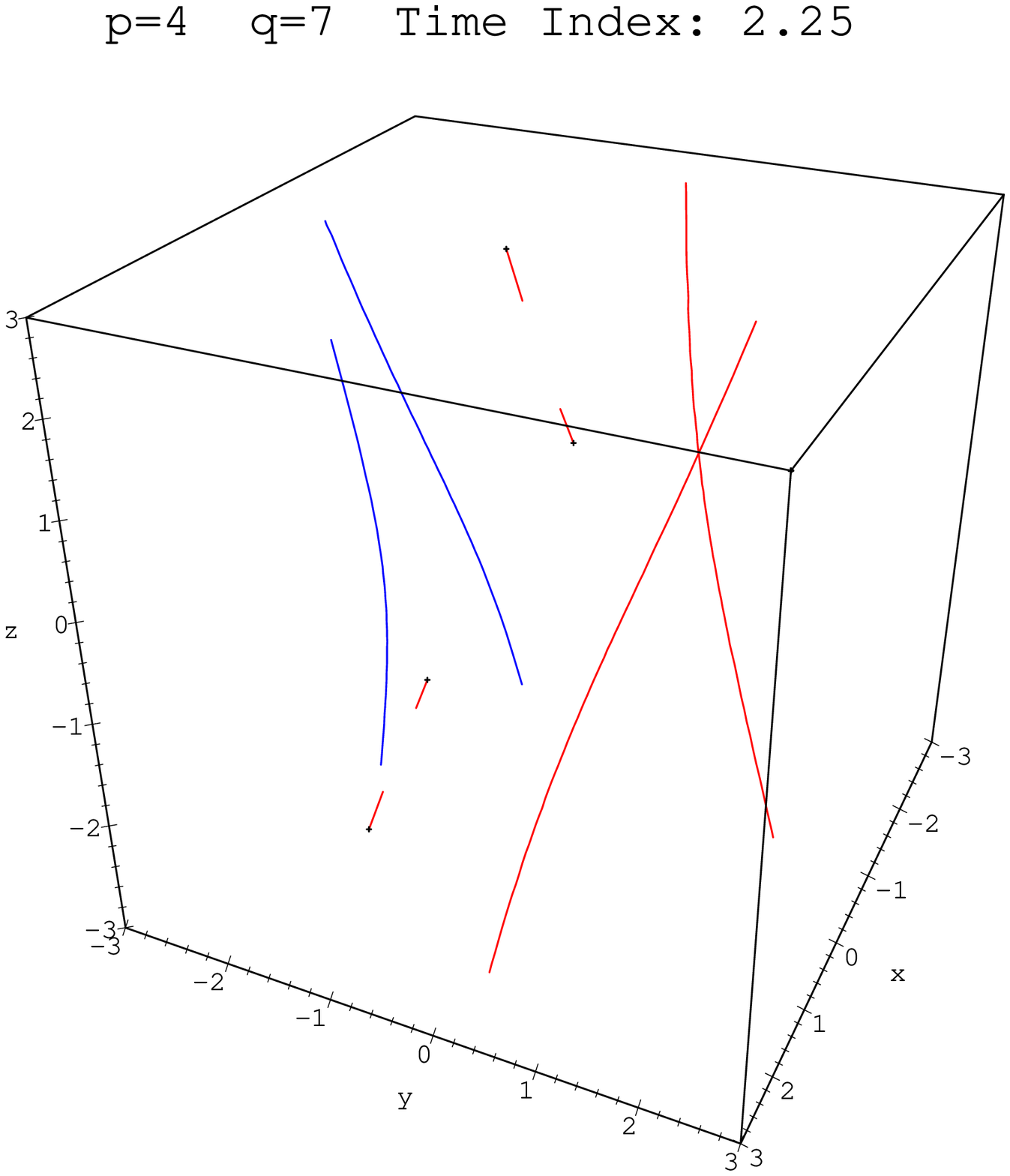,height=2.0in,width=2.0in}}
\put(390,200){\psfig{figure=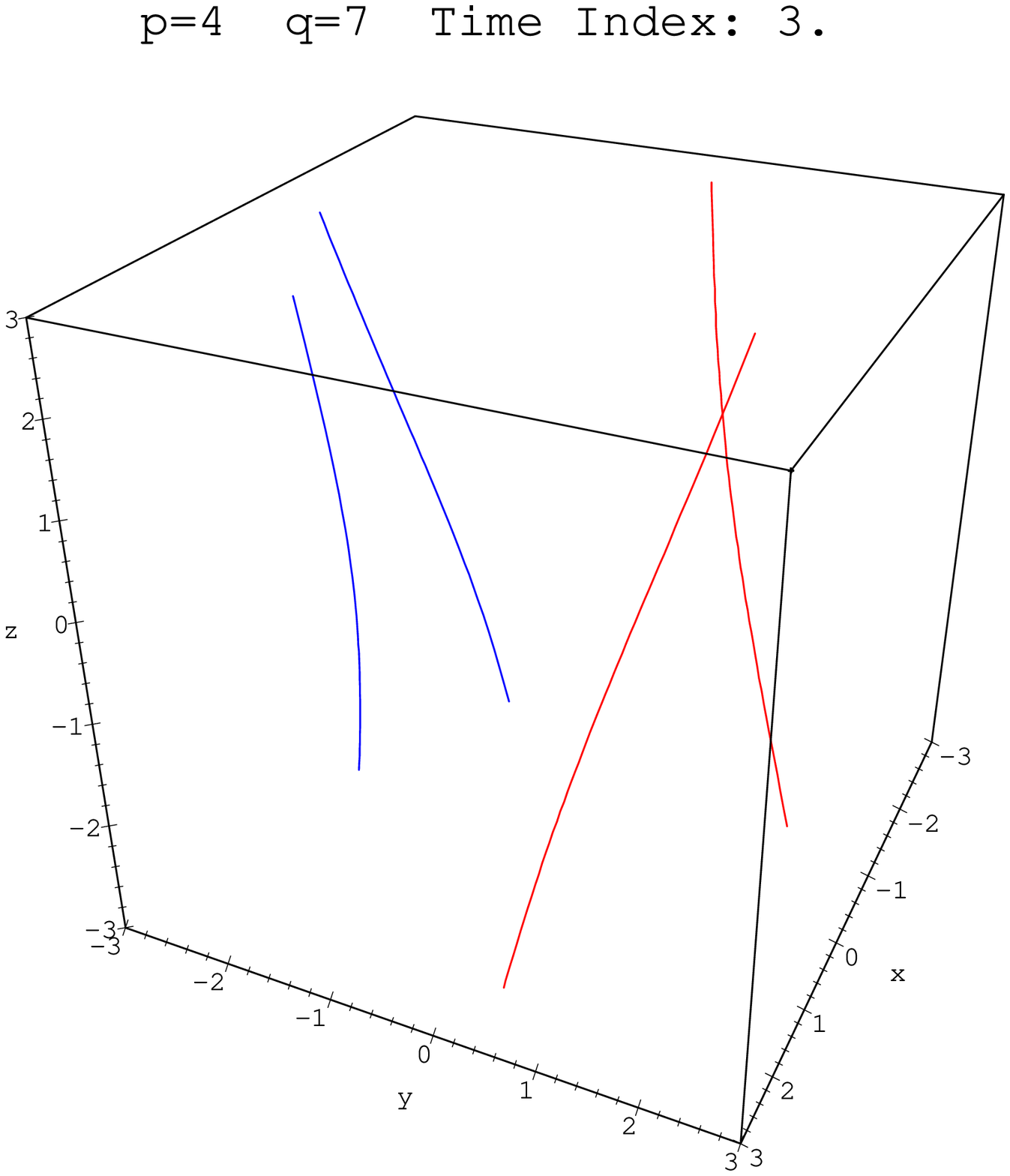,height=2.0in,width=2.0in}}
\end{picture}
\vskip6.0in
\caption[Evolution of the (4,7) torus link instanton.]
{Evolution of the (4,7) torus knot instanton.}
\end{figure}
\eject

Figure 5 shows a sequence of frames of the (4,3) torus knot instanton.
Initially, we see two string segments; at time $t \approx -1.0$, however,
a pair of points appear. As time continues to evolve these points develop
a string segment connecting the two. This finite segment interacts with
the pair of strings initially seen, then the two original segments
combine and disappear. This behavior is reminiscent of the early meson
string model, in which a meson was seen as a $q\bar{q}$ pair connected by a
string of gluon flux. In this interpretation, it appears as though a
$q\bar{q}$ pair is created out of the vacuum, interacts strongly with a pair
of gluon flux lines, then is annihilated back to the vacuum in the presence
of a monopole/anti-monopole pair.  Here it is the original 
string segment that we interpret as monopole-anti-monopole pairs. 
We emphasize that these are open string solutions that provide this
topology as opposed to closed stings. The new requirement is that the
opens strings be self intersecting.

The structure of the images suggests that the quarks interact with
each other through non-local interactions.  We would like for the
strings to correspond to flux tubes of gauge fields and to have quarks
on the endpoints.   Lets fix the gauge, $A_0=0$ and    
consider the Wilson line integral given by the following path-ordered
exponential,
\begin{equation} U({\bar X}(\sigma_{+},t), {\bar X}(\sigma_{-},t))= {\cal P}
\exp{(\int^{\sigma_+}_{\sigma_-} A_i {d X^i\over d \, \sigma} d\,\sigma)}.
\end{equation}
Here ${\bar X}(\sigma_{+})$ and ${\bar X}(\sigma_{-})$ are the end
points of the string as seen from a fixed frame and $\sigma$ is used
to parameterize the string for each time, $t$.  
From  the image of the string
immersions it appears that  non-local point
particle-like interaction Lagrangians such
\begin{equation}
 L={\bar q}({\bar X}(\sigma_{+},t))\,U({\bar X}(\sigma_{+},t), {\bar
      X}(\sigma_{-},t))\, q({\bar X}(\sigma_{-},t)). 
\end{equation} 
can be supported. 
In order to account for the semi-infinite segments
we can imagine taking one of the endpoints of the Wilson line out to
spatial infinity.  At spatial infinity the gauge fields approach zero
and the Wilson line goes to the identity.  The semi-infinite segment 
then can support a quark attached to the end of the segment with a
monopole or anti-monopole gauge field confined to the segment.  
One can build an operator that respect the residual gauge invariance with
\begin{equation}
 {\bar q}({\bar X}(\sigma_{+},t))\,U({\bar X}(\sigma_{+},t),
 \infty),\label{eq:wilsonline1} 
\end{equation}
or either 
\begin{equation}U(\infty, {\bar X}(\sigma_{-},t)) q({\bar
  X}(\sigma_{-},t)). \label{eq:wilsonline2}
\end{equation}
The lines that are infinite in extent correspond to both Wilson line
endpoints being taken to spatial infinity.  
These configurations seem to be the catalyst for 
pair production and annihilation.  
There might be suppression to these infinite and semi-infinite segments
if one uses  the same ``area law'' arguments
[\ref{bib:Wilson}] one uses to suppress largely separated quark paths
in pair production if the flux tubes are too far from each other.
However very close pairs ${\bar q}\,U$ and $U\, q$ may be significant 
and could contribute to $<{\bar q} q>$.
To sum up the overall picture that these string suggests in that 
the general vacuum structure will consist of infinite and semi-infinite
flux tubes as well as emerging and disappearing finite tubes and
stable finite strings.  The gauge fields will be confined to the
surfaces of these segments suggesting that monopole and anti-monopole
are pervasive and where pair production seems to be catalyzed by the
presence of these monopoles/anti-monopoles.  The presence of the flux
tubes pouring their gauge flux out at spatial infinity is 
already suggested by the 
electric Meissner effect [\ref{bib:Rosensweig}].  The
exchange of flux tubes between pair-produced segments and the other
segments suggest the charge-exchange scattering processes between
fermions and charged Dirac magnetic monopoles [\ref{bib:Christ}].

\section{${\bar q} q $ Production}
The topology that supports the pair production and annihilation is not
due directly to string self-intersection but
is related to how the time parameter $X^4$ takes its values from the 
$\sigma - \tau$ parameter space.  This is related to the fact that the
solution is self-dual, however.   
Let us
consider the multi-valued relationship  
of $\tau$ with $\sigma$ for a given
time,  $X_4$.  Since the $X_4$ coordinate only depends on $q$, we can
examine a time sequence of the relationship of $\sigma$ vs. $\tau$.  
To be explicit let us consider any solution where $q=3$.  
\begin{figure}[b]
\begin{picture}(100,200)(60,600)
\put(50,680){\psfig{figure=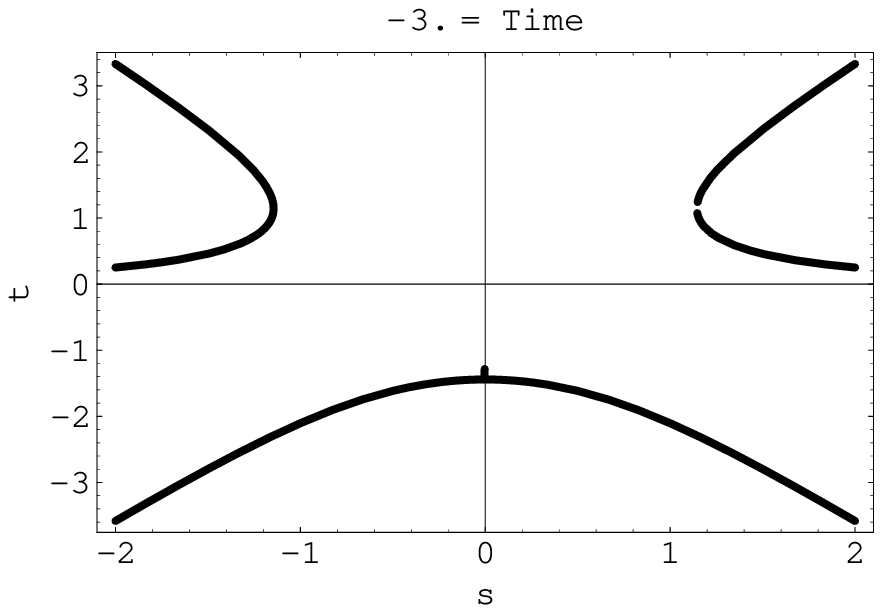,height=2.0in,width=2.0in}}
\put(220,680){\psfig{figure=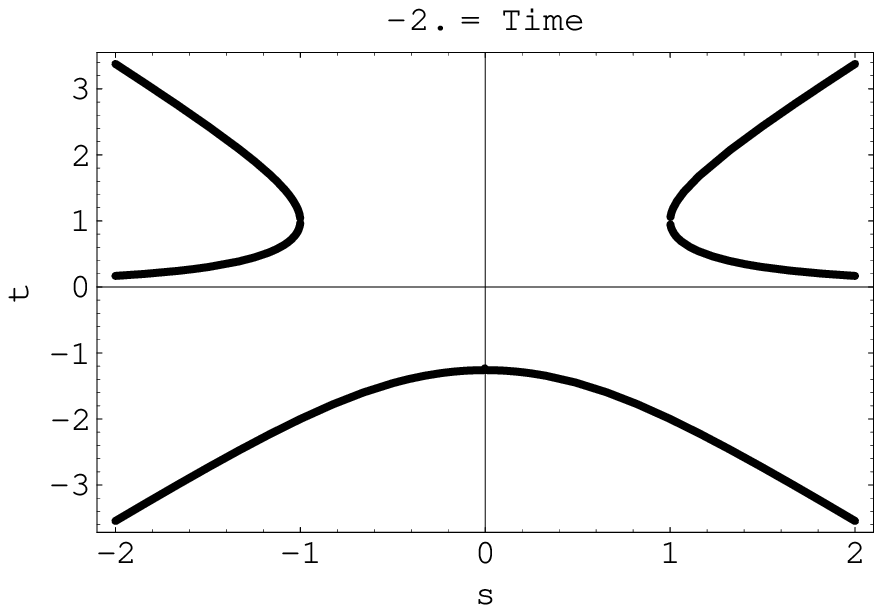,height=2.0in,width=2.0in}}
\put(390,680){\psfig{figure=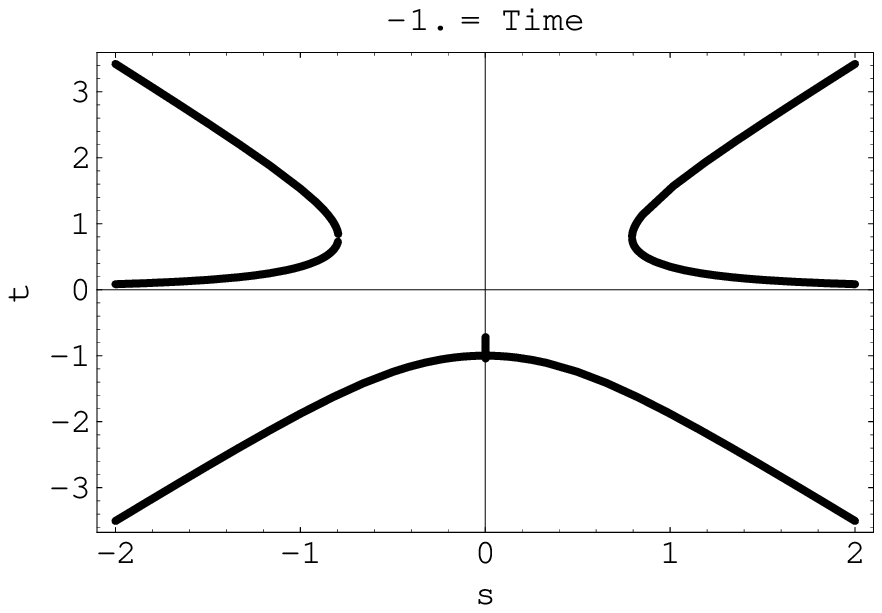,height=2.0in,width=2.0in}}
\put(50,530){\psfig{figure=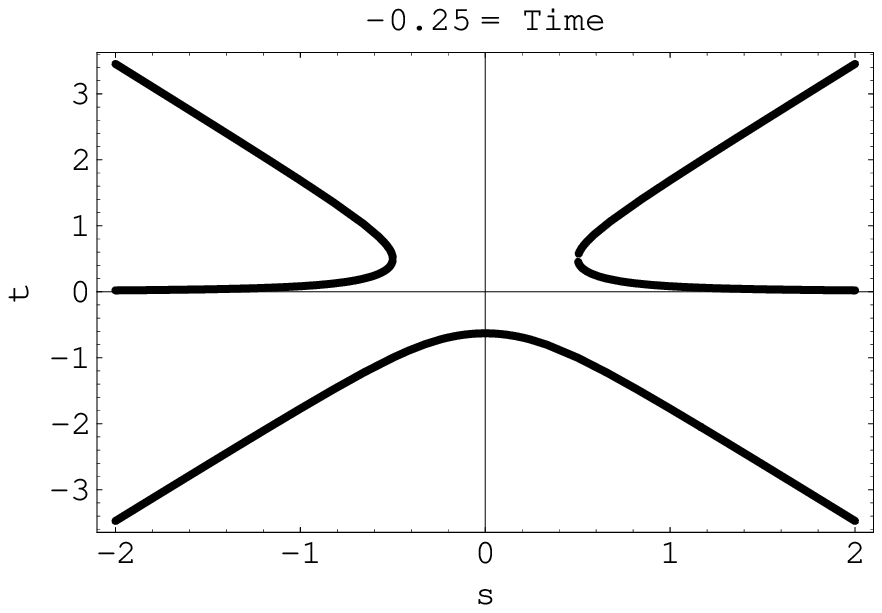,height=2.0in,width=2.0in}}
\put(220,530){\psfig{figure=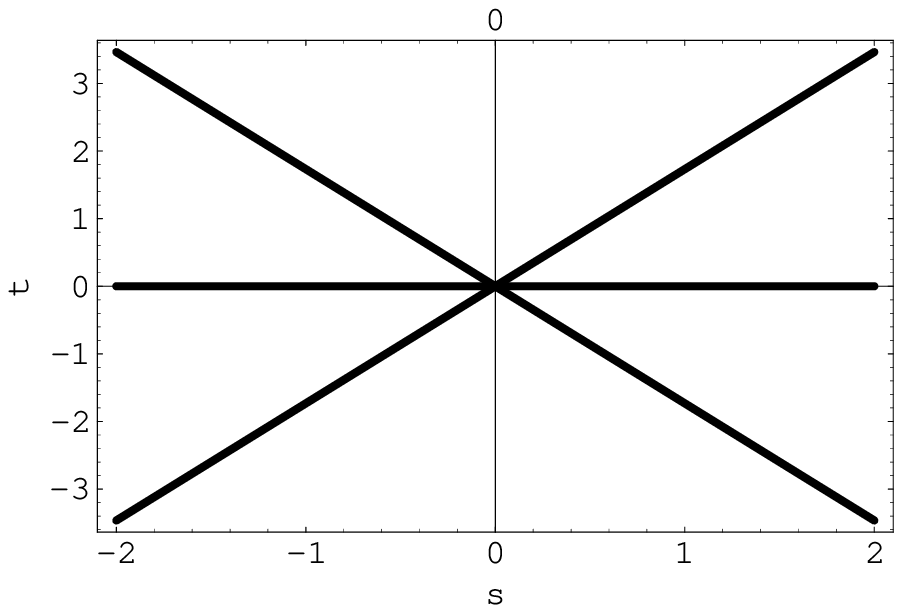,height=2.0in,width=2.0in}}
\put(390,530){\psfig{figure=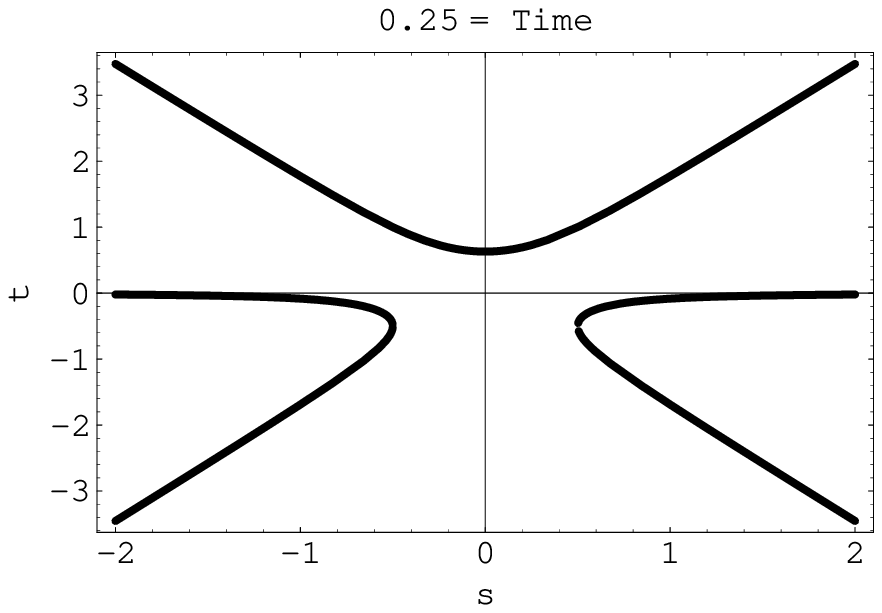,height=2.0in,width=2.0in}}
\put(50,380){\psfig{figure=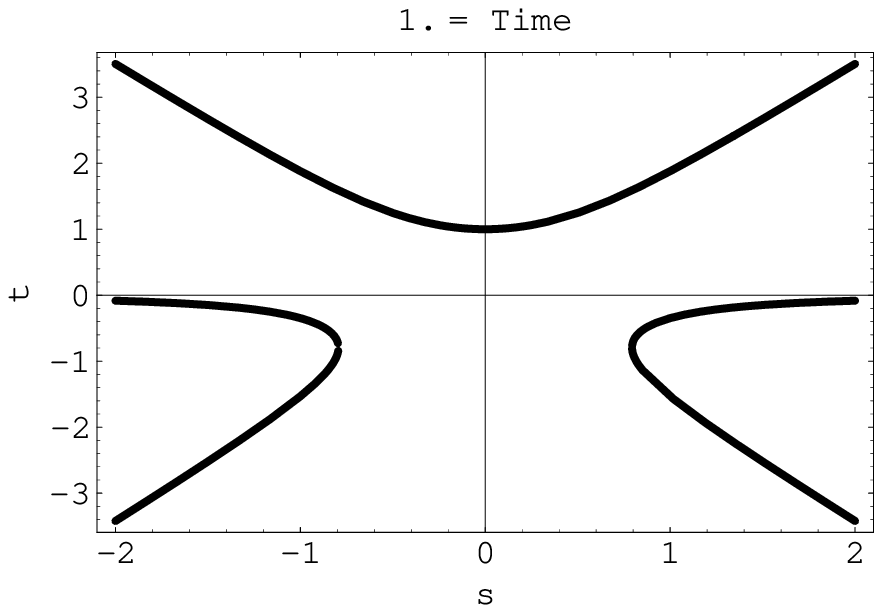,height=2.0in,width=2.0in}}
\put(220,380){\psfig{figure=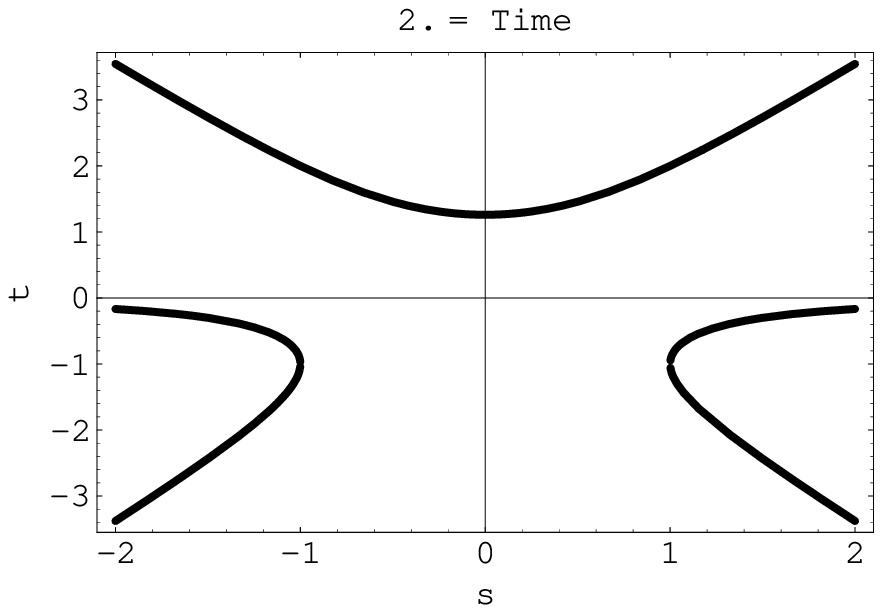,height=2.0in,width=2.0in}}
\put(390,380){\psfig{figure=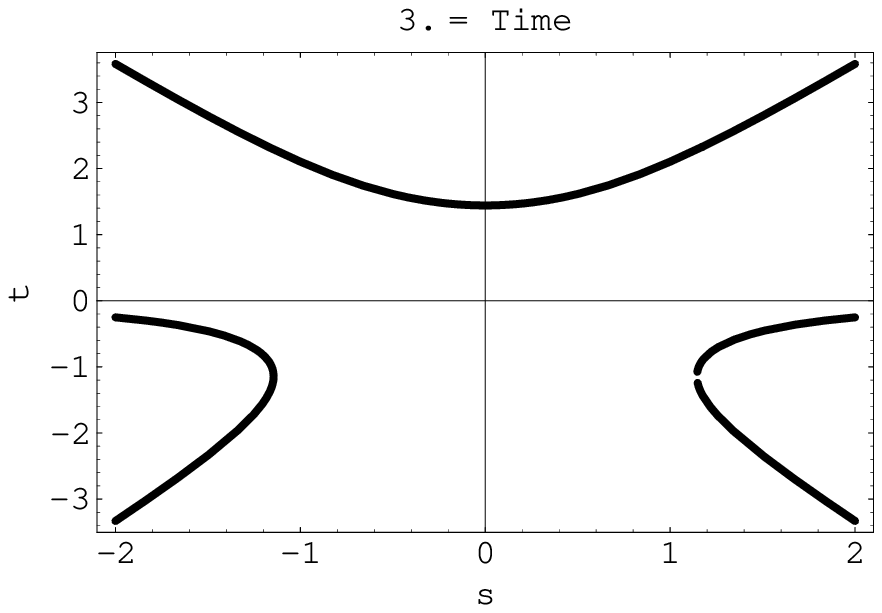,height=2.0in,width=2.0in}}
\end{picture}
\vskip3.0in
\caption[$\sigma$ vs. $\tau$ at different times]{$\sigma$ vs. $\tau$
  at different times for $q=3$} \label{fig:production}
\end{figure}
Here one sees that at very early times ($t<< -1$), $\tau$ is a single
valued function for values of $\sigma$ between $-1$ and $1$.  As time 
moves on, this function becomes multi-valued (here at $t=-2$) 
for values where $\sigma$ takes values from -1 to 1. Now  the physical 
$\sigma-\tau$ parameter space is split into three distinct regions
corresponding to three distinct images in the frame.  
As time moves on these region move toward the origin until at $t=0$
all
three regions merge for an ``instant'' and the regions begin to recede.
At some later time (here $t=2$) the $\tau$
parameter is again single valued with respect to the physically
relevant values for $\sigma$.   This picture highlights two issues
about the self-dual solution that may be realized in  the strong
coupling regime of QCD.  The first is how the multivalued nature of $\tau
$ vs. $\sigma$  can cause pair production and annihilation and the
second being the singularity of the map at $t=0$, the point of
self-intersection, leads to regions where chiral symmetry can be
broken.  These singular points are where the string segments are
allowed to exhange flux lines and also where the determinant of the
induced metric vanishes.  The details of this last remark will be
explored in the next section.  For now we will take a closer look at
the issues related to production and annihilation.  

From figure [\ref{fig:production}], one sees that from the time of the
production ($t=-2$)  until the time of annihilation ($t=2)$, four  units
of time have transpired.  Here we can assume that the constant $\mu$
from the action in Eq.[1],
sets the space and time scale since $\mu$ has dimensions of
$[L^{-2}]$.   From the uncertainty principle we can estimate a typical
value of $\frac 1{\sqrt{\mu}}$ since four quarks are produced in a
time $4  \frac 1{\sqrt{\mu}}$.  This would imply that $\mu \approx (16
M_{q})^2$ where $M_q$ is the quark mass.  The $q=3$ set of torus knots
correspond to the minimal pair-producing configuration.  
However higher $q$ instantons could easily surpass the bounds from
the uncertainty principle by producing more quarks in nearly
equal times to that of the $q=3$ solutions.  For this reason we need to check
the production capabilities of the $q>3$ solutions.  

In order to determine the time it takes for higher $q$ configurations
to produce and annihilate quarks we need only to ask at what time,
$t_{real}$ are all the roots of $\tau(\sigma=-1, t_{real})$ are real.  
For every value of $q$ there are $q$ 
regions that must merge into the $(-1 < \sigma < 1)$ region and then move
out again. The total production/annihilation time would then be $2
\times t_{real}$.  As an example lets examine the $q=4$ case.
\begin{figure}[b]
\begin{picture}(100,200)(60,620)
\put(50,680){\psfig{figure=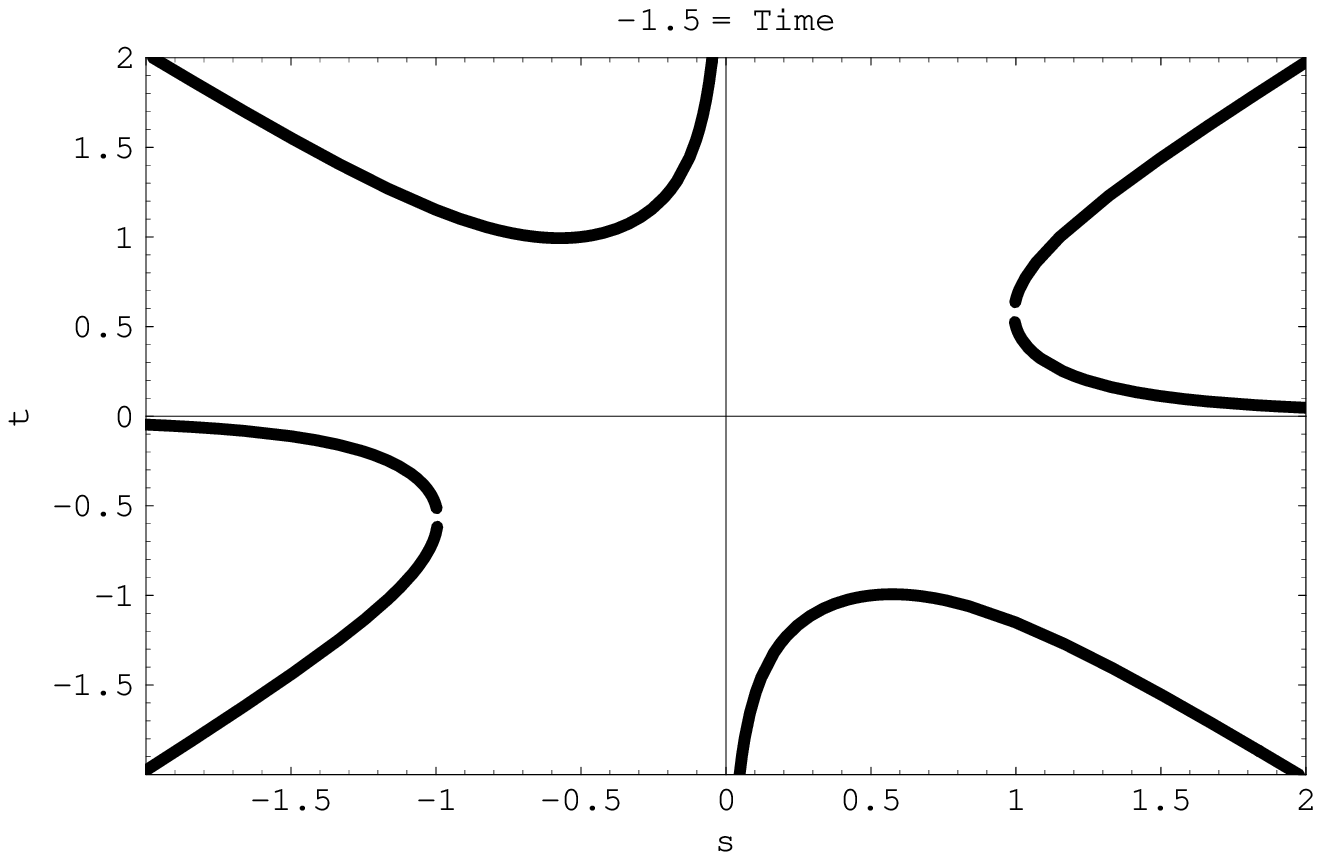,height=2.0in,width=2.0in}}
\put(220,680){\psfig{figure=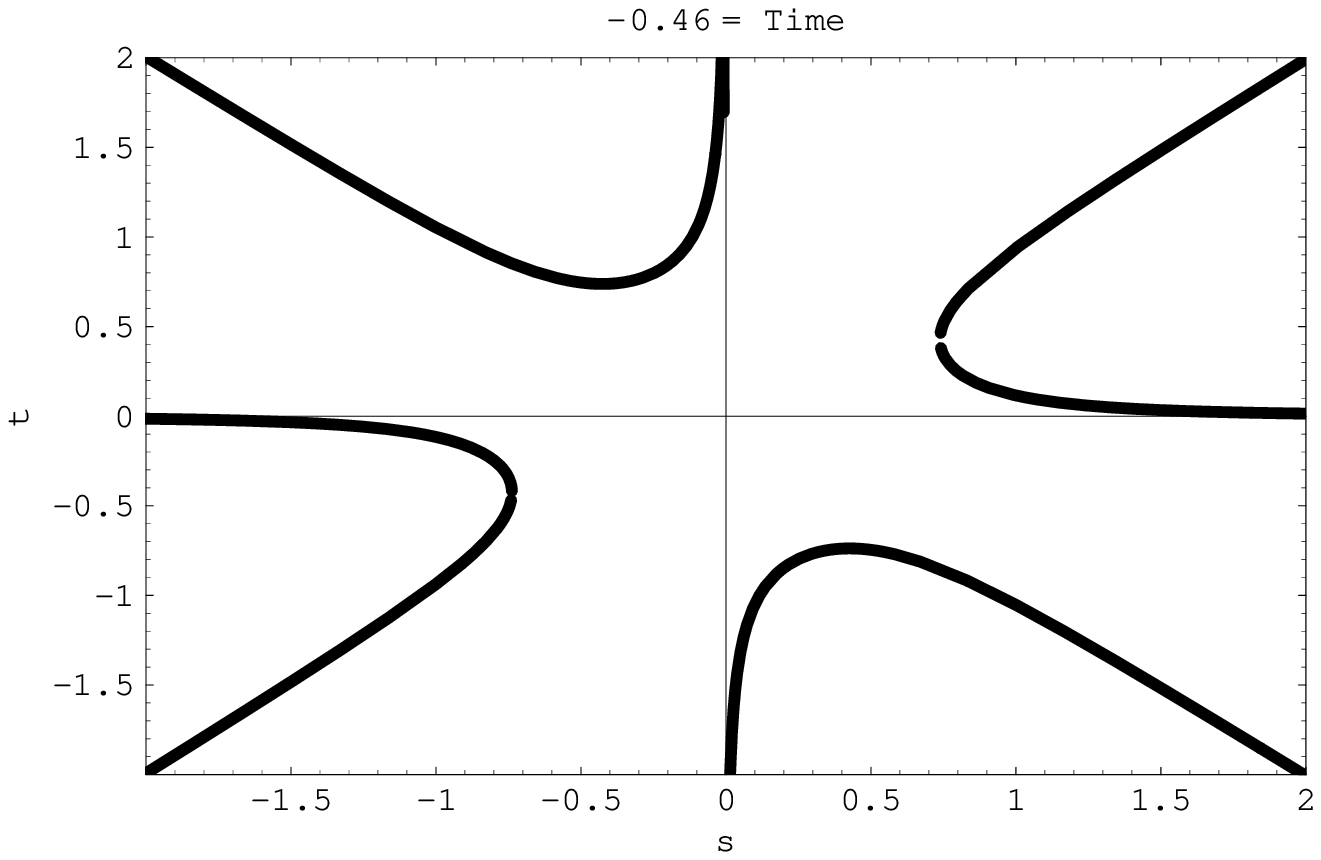,height=2.0in,width=2.0in}}
\put(390,680){\psfig{figure=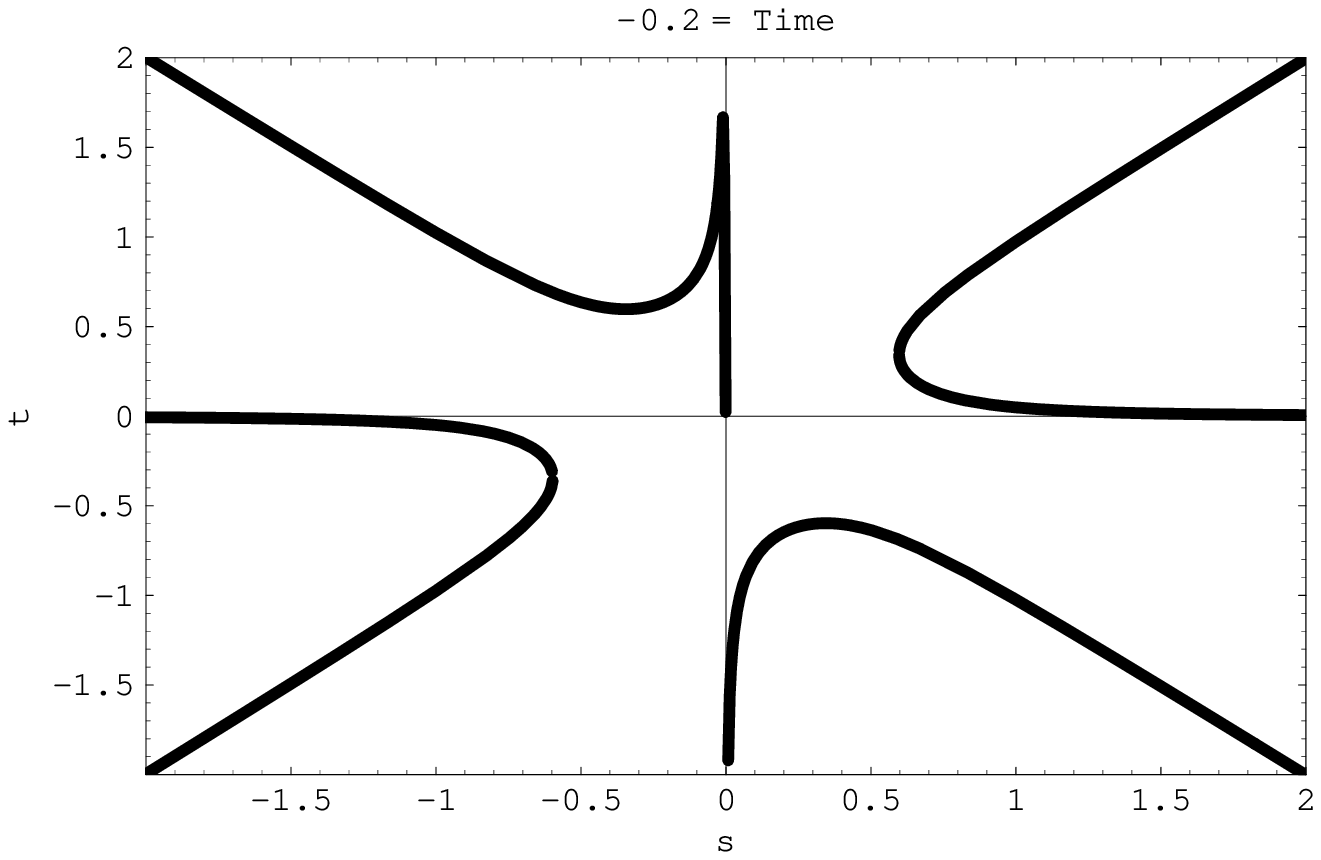,height=2.0in,width=2.0in}}
\put(50,530){\psfig{figure=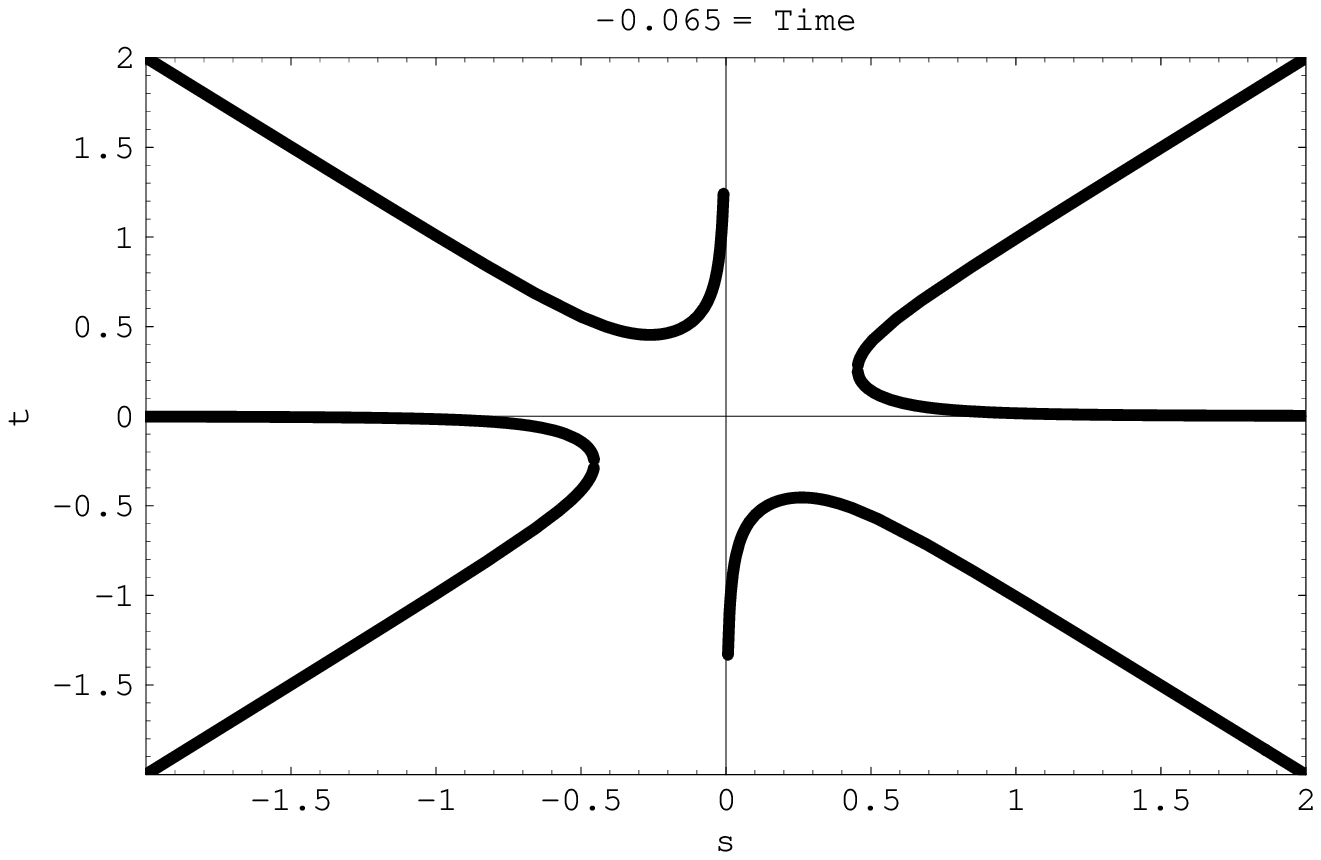,height=2.0in,width=2.0in}}
\put(220,530){\psfig{figure=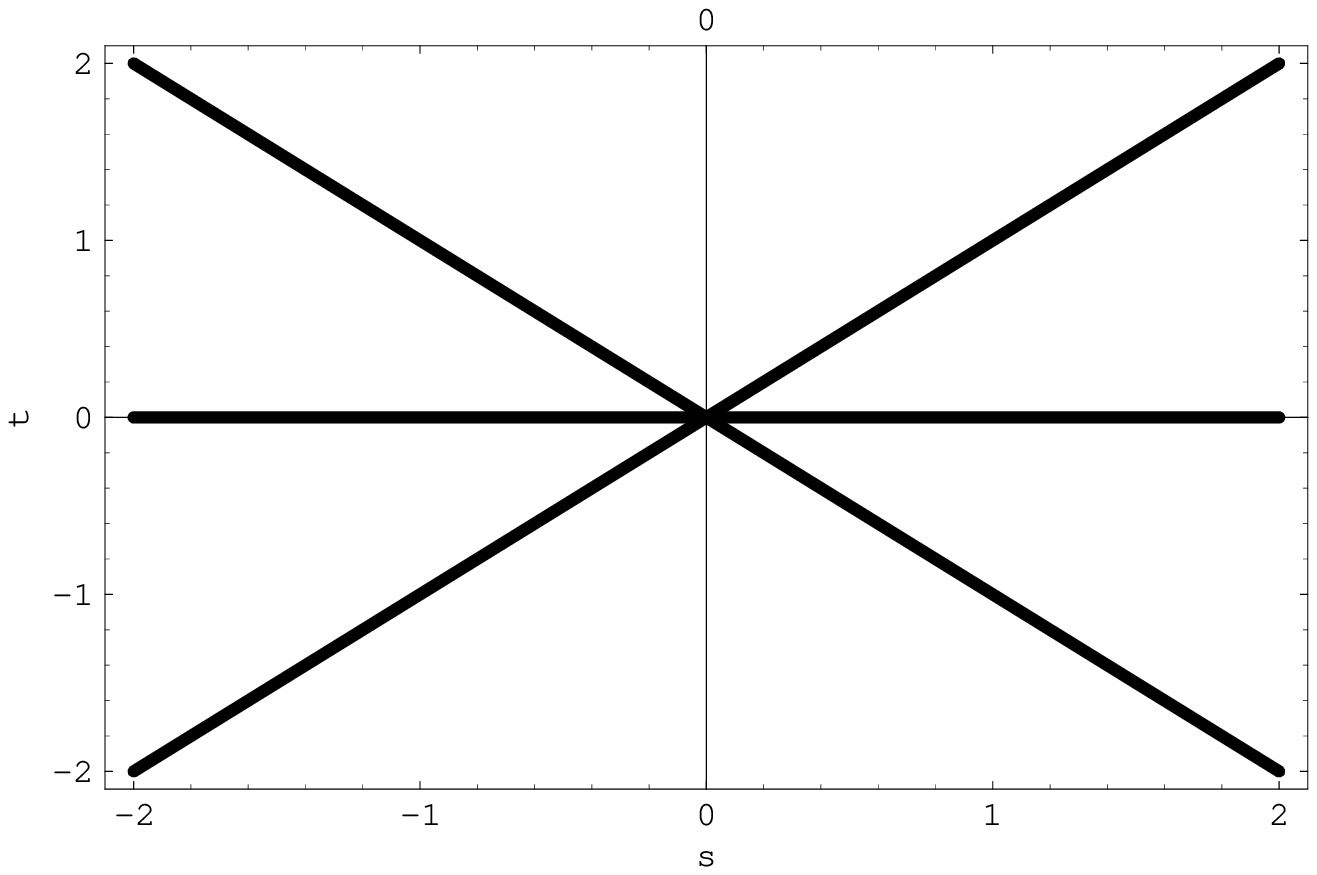,height=2.0in,width=2.0in}}
\put(390,530){\psfig{figure=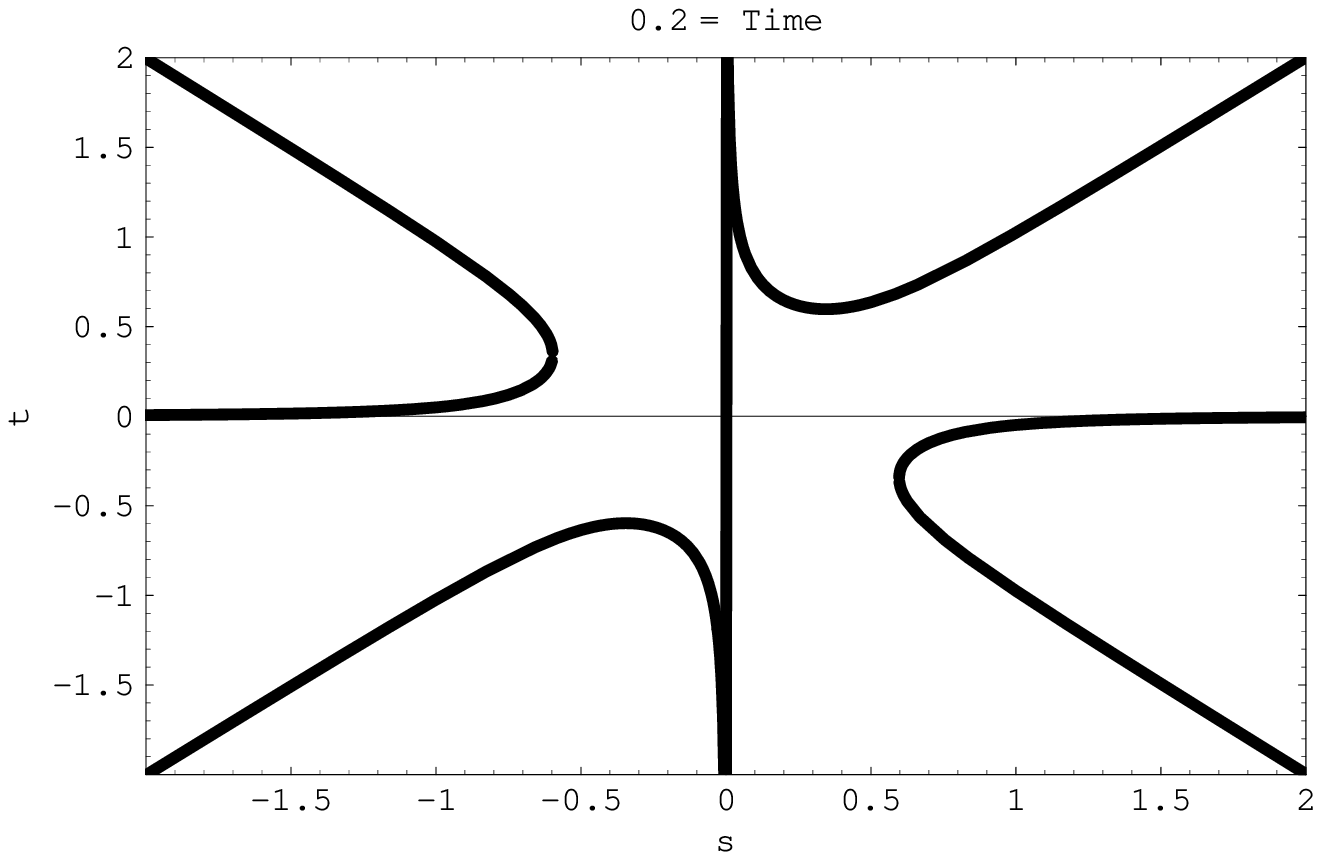,height=2.0in,width=2.0in}}
\put(50,380){\psfig{figure=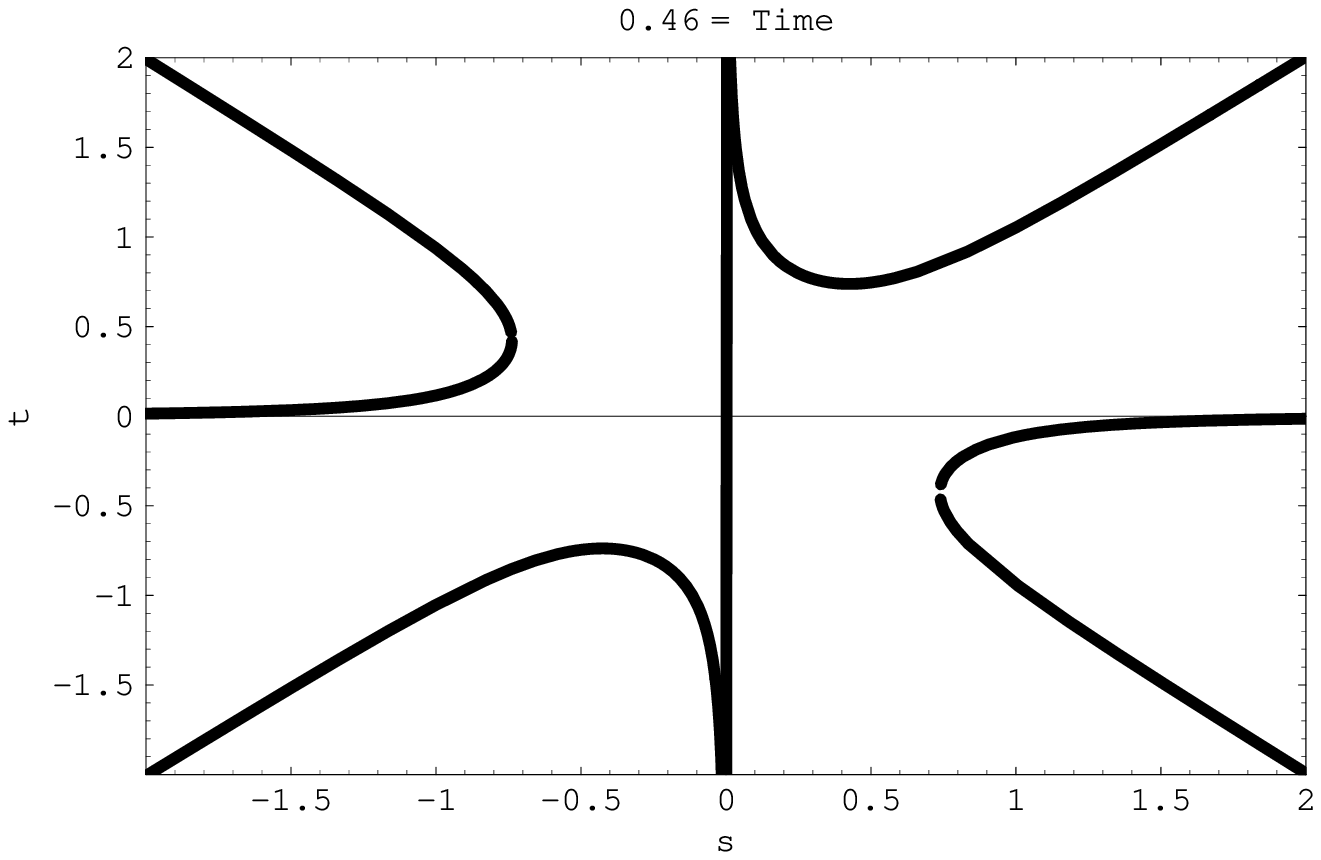,height=2.0in,width=2.0in}}
\put(220,380){\psfig{figure=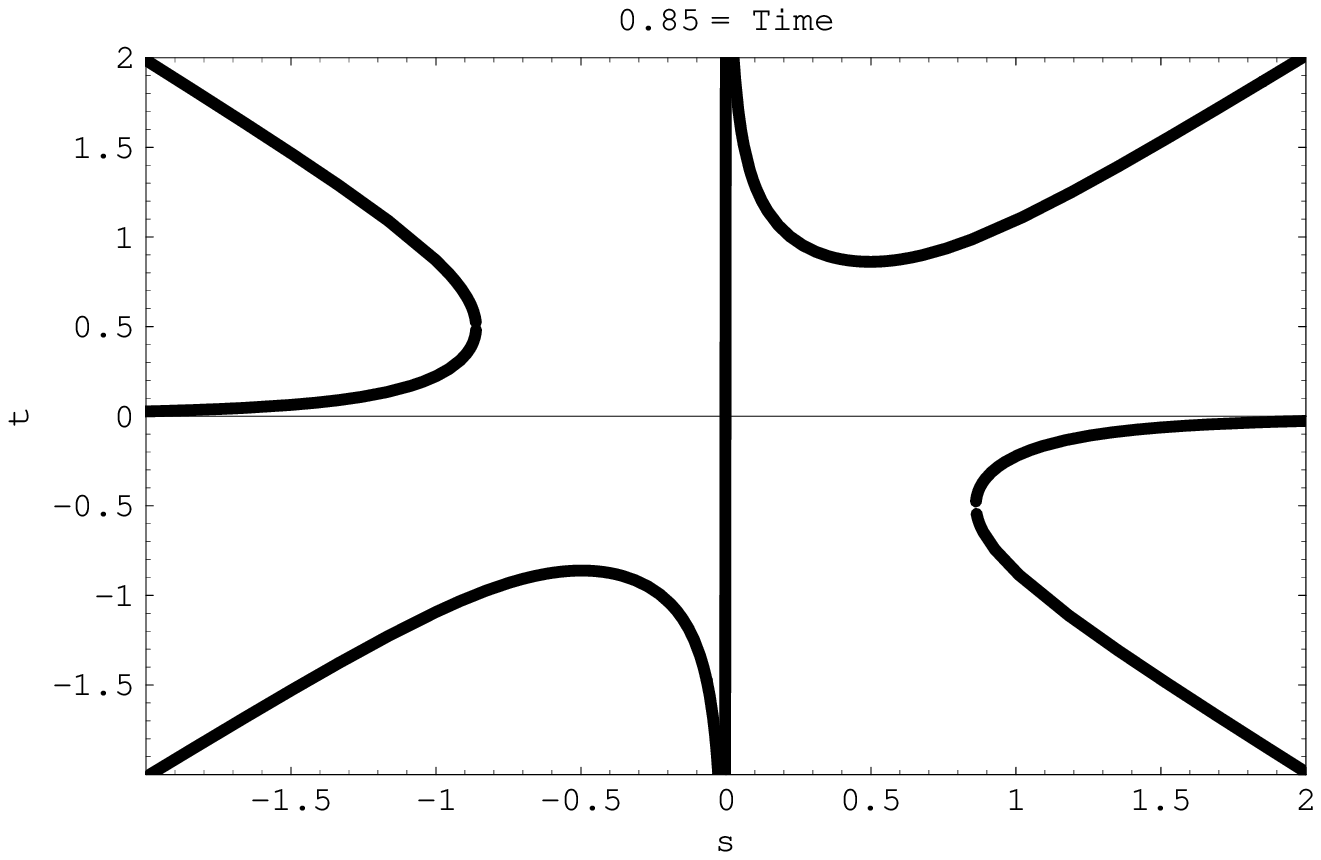,height=2.0in,width=2.0in}}
\put(390,380){\psfig{figure=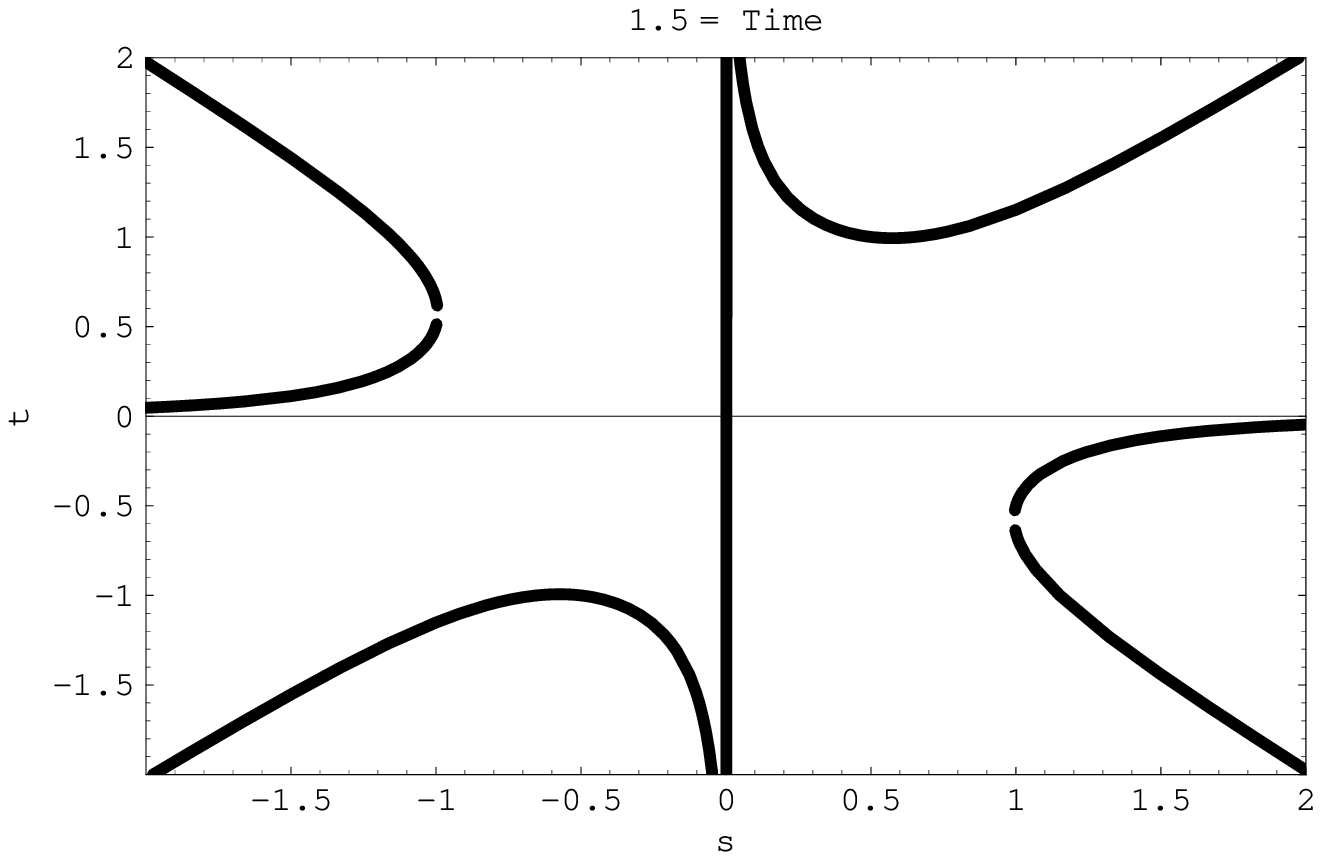,height=2.0in,width=2.0in}}
\end{picture}
\vskip3.5in
\caption[$\sigma$ vs. $\tau$ at different times]{$\sigma$ vs. $\tau$
  at different times} \label{fig:production2}
\end{figure}
In this case the pair production begins at $t= -\frac 8 {3 \sqrt{3}}$
where $\tau$ takes values of $\frac 2{\sqrt{3}}$,  $\frac{-1}{\sqrt{3}}$
 and $\frac {-1}{\sqrt{3}}$.  The time for total pair production is then
$\frac {16}{3 \sqrt{3}}$ time units or about three units of time.  
In this amount of time again four quarks or anti-quarks are produced
and annihilated.  As $q$ increases it empirically appears that the 
amount of time to undergo full pair production monotonically decreases
but reaches a plateau. One can show that the  total production times 
for different values of $q$ are:
\begin{eqnarray}
q=3, & \,\ \,\, \,\,t_{real}= 2 \\
q=4, &\,\, \,\, \,\, t_{real}=\frac 8{3 \sqrt{3}}  \\
q=5, &\,\, \,\, \,\, t_{real}=1.37 \\
q=6, &\,\,  \,\,\,\, t_{real}=1.285 \\
q=10, &\,\,  \,\,\,\, t_{real}=1.147 \\
q=100, &\,\, \,\, \,\, t_{real}=1.012 \\
q=111, &\,\, \,\, \,\, t_{real}= .608
\end{eqnarray}
We observe that $q=111$ puts an upper limit
on the value of $q$ that will produce any new pairs.  For all values
greater that $q=111$ some of the roots for $\tau $ remain complex even
when $t = 0$.  By analogy with the $q=3, 4$ and $5$ cases we see that 
there can be at most 112 quarks/anti-quarks produced in $1.2$ 
seconds by any instanton.   Since the total number of quarks is
limited by all torus knot instantons we can again use the uncertainty
principle to get an even better estimate of $\mu$.  We find that 
$\mu \approx (112 M_q)^2$.  

Animations of several $(p,q)$ solutions with $X^4$ chosen as the time
parameter can be found at 
http://www-hep.physics.uiowa.edu/\~{}bacus/research.html.

\section{Chiral Symmetry Breaking Processes}

As we discussed earlier, the self intersection of the world sheet can
cause singular points to exist and lead to zeros in the  determinant of the
induced metric $g_{a b} = \partial_a X^\nu \partial_b X_\nu$. Self
duality for the string instantons imply that $t^{\mu \nu}$ from
Eq.(\ref{eq:dual}) satisfy the constraint 
\begin{equation}
t^{\mu \nu} = \frac 12 \epsilon^{\mu \nu \lambda \rho} t_{\lambda
  \rho} + C^{\mu \nu},
\end{equation}
where $C^{1,2}=-C^{2,1}=C^{3,4}=-C^{4,3}=1$ and all other entries are
zero.
 
As we are interested in chiral symmetry, we would like to look at the
pullback of $\gamma_5$ onto the string world sheet.   In terms of
functions on the world sheet we may write 
\begin{equation}
{\tilde  \gamma_5} = {\epsilon^{c d} \over 4 \sqrt{g}} {\hat
  \epsilon^{E F}} \partial_c X^\mu \partial_d X^\nu N^\lambda_E
  N^\rho_F \frac 14 [\gamma_\mu, \gamma_\nu] \, [\gamma_\lambda,
  \gamma_\rho].
\end{equation}
In the above $N^\mu_A$ are elements of the normal bundle of the string
world sheet and ${\hat \epsilon}^{A B}$ is the volume form associated
with the normal bundle.  By using self duality we can eliminate the
tangent vectors from the above expression and write,
\begin{equation}
{\tilde \gamma_5} = \frac 14{\hat \epsilon}^{E F} C^{\mu \nu}
N^\lambda_E N^\rho_F \epsilon_{\mu \nu \lambda \rho} \gamma_5.
\end{equation}
The normal vectors satisfy the conditions that 
$\partial_a X^\mu N^\mu_E =0,$ and $N^\mu_E N^{\mu\,F} = \delta^F_E,$.
These conditions as well as the self-duality conditions imply that a 
solution for the tangent and normal vectors is 
\begin{eqnarray}
\partial_\sigma X^\mu &= \big(\partial_\sigma X^1, \partial_\sigma X^2, 
\partial_\sigma X^3, \partial_\sigma X^4\big)\\  
\partial_\tau X^\mu &= \big(-\partial_\sigma X^2, \partial_\sigma X^1, 
\partial_\sigma X^4, -\partial_\sigma X^3\big)\\ 
N^\mu_1 &=\frac 1{\sqrt{g}} \big(\partial_\sigma X^3, \partial_\sigma X^4, 
-\partial_\sigma X^1, -\partial_\sigma X^2\big)\\  
N^\mu_2 &=\frac 1{\sqrt{g}} \big(\partial_\sigma X^4, -\partial_\sigma X^3, 
\partial_\sigma X^2, -\partial_\sigma X^1\big), \label{eq:bundle}
\end{eqnarray}
where $g$ is the determinant of the induced metric.  For these
solutions $\sqrt{g} = g_{\sigma \sigma}$. Without loss of generality we
can take the (3,2) torus  knot as an example. Here the 
torus knot determinant is 
\[ \sqrt{g} = 9(\tau^4 + \sigma^4) -12 \tau^2 \sigma^2 + 4 \sigma^2 +
4 \tau^2. \]
Clearly this vanishes at $\sigma = 0, \tau = 0$.  
This corresponds to the point where the string self-intersects.  At
this point the normal bundle is ill-defined and two copies of the
tangent vectors are imaged at this point.  
The pullback of $\gamma_5$ is also ill-defined when the determinant
vanishes.
 Elsewhere on the manifold 
${\tilde \gamma_5}$ can be used to construct a projection operator for
chiral symmetry.  However these points where the determinant vanish 
 act as tiny ``bubbles'' where chiral symmetry
is broken.  The transition of the normal bundle going through this
singularity has the effect of imposing reflections on some but not
necessarily all of the
components. In the (3,2) case the relationship of the normal vectors
just before and just after is given by:
\begin{equation}
{N^\mu_1}_{before} =\frac 1{\sqrt{g}} \big(\partial_\sigma X^3, \partial_\sigma X^4, 
-\partial_\sigma X^1, -\partial_\sigma X^2\big)
\end{equation}
while 
 \begin{equation}
{N^\mu_1}_{after} =\frac 1{\sqrt{g}} \big(\partial_\sigma X^3,- \partial_\sigma X^4, 
-\partial_\sigma X^1, \partial_\sigma X^2\big)
\end{equation}
 flipping the second and fourth components.  Similar changes happen in
 both the tangent vectors and the other normal.  
To show that the orientation has changed consider the vector 
${\bar M} = {\bar N_1} \times {\bar N_2}$ for the (3,2) knot. 
Just after the singularity, $M_y \to -M_y$ while all other components
 remain the same.  This corresponds to a change in the
 orientation.  This pictures corresponds to the field theoretic case
 where the core of monopoles can be thought of as bubbles where chiral
 symmetry is broken.  The S-wave of the fermions interact with this
 core which leads to processes such as those seen in [\ref{bib:Christ},\ref{bib:Callan}].

\section{ The Self-Intersection Number}
It is conjectured in [\ref{bib:Robertson}] that the self-intersection
number of the $(p,q)$ torus is $\nu = 4(p-q)$.
Here we  explicitly calculate the self-intersection number $\nu$ 
for a general (p,q) torus-knot solution proving the
conjecture. Consider an open string where $(-L \le \sigma \le L)$ and
$(-\infty < \tau < \infty)$.  Starting from the general definition of the
self-intersection number
\begin{eqnarray*}
\nu = \frac{1}{2 \pi} \int_{-\infty}^{\infty} d\tau \int_{-L}^{L} d\sigma
\partial_{a} t^{\mu \nu} \partial_{a} t^{\mu \nu}
\end{eqnarray*}
which we can express in terms of the string solution $X(\sigma, \tau)$ as
\begin{eqnarray*}
\nu = \frac{1}{2 \pi} \int_{-\infty}^{\infty} d\tau \int_{-L}^{L} d\sigma
\partial_{a} (\frac{\epsilon^{c d}}{\sqrt{g}}
\partial_c X^{\mu} \partial_d X^{\nu}) \partial_{a} (\frac{\epsilon^{c d}}
{\sqrt{g}} \partial_c X^{\mu} \partial_d X^{\nu})
\end{eqnarray*}
Using the general (p,q) torus-knot solution
\begin{eqnarray*}
X(\sigma, \tau) = [(\sigma^2 + \tau^2)^{p/2} \sin(p \tan^{-1}(\frac{\sigma}
{\tau})), (\sigma^2 + \tau^2)^{p/2} \cos(p \tan^{-1}(\frac{\sigma}{\tau})), \\
-(\sigma^2 + \tau^2)^{q/2} \cos(q \tan^{-1}(\frac{\sigma}{\tau})),
-(\sigma^2 + \tau^2)^{q/2} \sin(q \tan^{-1}(\frac{\sigma}{\tau}))]
\end{eqnarray*}
we find
\begin{eqnarray*}
\nu = \frac{1}{2 \pi} \int_{-\infty}^{\infty} d\tau \int_{-L}^{L} d\sigma
\frac{8 p^2 q^2 (p-q)^2 (\sigma^2 + \tau^2)^{p+q-1}}
{(p^2 (\sigma^2 + \tau^2)^p + q^2 (\sigma^2 + \tau^2)^q)^2}
\end{eqnarray*}
If we let $r^2 \equiv \sigma^2 + \tau^2$,
\begin{eqnarray*}
\nu &=& \frac{1}{ \pi} \int_{-\pi /2}^{\pi /2} d\theta \int_{0}^{\frac{L}
{\cos(\theta)}} dr
\frac{8 p^2 q^2 (p-q)^2 r^{2(p+q-1)+1}}
{(p^2 r^{2p} + q^2 r^{2q})^2} \\
 &=& \frac{4((p-q) q p)^2}{\pi} \int_{-\pi /2}^{\pi /2}  d\theta
\frac{1}{(p-q)q^2 (q^2 r^{2(p-q)} + p^2)} \left|_{0}^{\frac{L}
{\cos(\theta)}} \right.
\end{eqnarray*}
Let $\phi \equiv 2 \theta + \pi$,
\begin{eqnarray*}
\nu = \frac{2(p-q)}{\pi} \int_0^{2 \pi} \frac{1}{
(\frac{p}{q} L^{(p-q)})^2(\frac{1-\cos(\phi)}{2})^{(p-q)} + 1} \,d\phi - 2(p-q)
\end{eqnarray*}
Note the following relation;
\begin{eqnarray*}
\int_0^{2 \pi} f(\sin \theta, \cos \theta) d\theta =
-i \oint_{\frac{unit}{circle}} f(\frac{z-z^{-1}}{2i}, \frac{z+z^{-1}}{2}) \frac{dz}{z}
\end{eqnarray*}
Then
\begin{eqnarray*}
\nu &=& -i \frac{2(p-q)}{\pi} \oint_{\frac{unit}{circle}} \frac{dz}
{z((\frac{p}{q} L^{(p-q)})^2 (\frac{1}{2}(1- \frac{z+z^{-1}}{2}))^{(p-q)} +1)
}
-2(p-q) \\
 &=& 4(q-p) - \frac{4(q-p)}{\pi}  \pi \sum \mbox{Residues in unit circle} \\
 &=& 4(q-p) (1- \sum \mbox{Residues in unit circle})
\end{eqnarray*}
Now we must find the singular points of $f(z) = \frac{1}{z((\frac{q}{p}
{L}^{p-q})^2 (\frac{1}{2}(1- \frac{z+z^{-1}}{2})^{p-q} +1)}$; they are
\begin{eqnarray*}
z &=& 0 \\
z &=& 1-2(-(\frac{q}{p})^2(L)^{2(p-q)})^{\frac{1}{q-p}} \pm
2 \sqrt{(-
(\frac{q}{p})^2(L)^{2(p-q)})^{\frac{1}{q-p}} +
(\frac{q}{p}(L)^{(p-q)})^{\frac{4}{q-p}}}
\end{eqnarray*}
Evaluating the residue $(z-z_0) f(z)|_{z=z_0}$ gives zero; then we have
\begin{eqnarray*}
\nu = 4(q-p)
\end{eqnarray*}
for (p,q) torus-knot solutions. This confirms the conjecture of
Robertson [\ref{bib:Robertson}].

\section{Computer Program}
The software described in this section was used to implement and generate
the animated solutions and to perform some lengthy calculations. Two
symbolic computation systems were used; \it Maple \rm and \it Mathematica\rm.

Listing 1 shows the \it Maple \rm source code which generates raw ($x, y, z$)
coordinates for a (p,q) torus knot solution and exports them to a file. The
user selects the values for $p$ and $q$, the frame numbers to start and end
at, the number of data points to use for each frame, and the ranges for the
parameters $t$ and $\sigma$.

\renewcommand{\baselinestretch}{.5}
\subsubsection*{Listing 1}
\tt
\#----------------------------------------------------------------------\# \\
\# Animation of the string instanton solutions. This program generates a\# \\
\# .dat numeric data file for each frame of the animation sequence.     \# \\
\# Usage (on UNIX): maple 4D\_TorusKnot\_Data.txt \&                    \# \\
\#                                                                      \# \\
\# Bob Bacus c.1997                                                     \# \\
\#------------------------- User input parameters ----------------------\# \\
p:=3:                 \# F=z\^ p                                          \# \\
q:=2:                 \# G=-z\^ q                                         \# \\
startframe:=0:        \# Frame to start from                            \# \\
endframe:=400:        \# Frame to end on                                \# \\
points:=400:          \# Number of data points per frame                \# \\
lmin:=-3:lmax:=3:     \# Range for the time parameter                   \# \\
smin:=-1:smax:=1:     \# Range for the length parameter                 \# \\
\# ---------------------------- Main Program ---------------------------\# \\
gc(300000):interface(screenwidth=500):Digits:=4:                         \\
readlib(unassign):readlib(write):                                        \\
unassign('t'):unassign('l'):unassign('n'):unassign('u'):unassign('s'):   \\
assume(t,real):assume(s,real):                                           \\
z:=t+I*s:                                                                \\
F:=expand(z\^p):G:=expand(-z\^q):                                           \\
imf:=Im(F):ref:=Re(F):reg:=Re(G):img:=Im(G): \# X=[Im(F),Re(F),Re(G),Im(G)]  \\
l:=(lmax-lmin)/(endframe-startframe)*n+lmin: \# n represents frame number; l is the time\\
v:=(smax-smin)/(points-1):                                                  \\
for n from startframe by 1 to endframe do                                   \\
\hspace*{.2in}open(cat(convert(p,string),`x`,convert(q,string),`-`,convert(n,string),`.dat`)):\\
\hspace*{.2in}writeln(evalf(l)):                                                        \\
\hspace*{.2in}writeln(points):                                                          \\
\hspace*{.2in}for i from 1 to q-1 do                                                    \\
\hspace*{.2in}writeln(cat(`Solution \#`,convert(i,string))):                           \\
\hspace*{.4in}for u from smin by v to smax do                                         \\
\hspace*{.6in}r:=[fsolve(subs(s=u,img)=l,t,complex)]:r:=r[i]:                       \\
\hspace*{.6in}X:=expand(subs(t=r,imf)):Y:=expand(subs(t=r,ref)):Z:=expand(subs(t=r,reg)):\\
\hspace*{.6in}writeln(evalf(subs(s=u,[X,Y,Z]))):                                    \\
\hspace*{.4in}\hspace{.2in}od;                                                                     \\
\hspace*{.2in}od;                                                                       \\
\hspace*{.2in}close():                                                                  \\
od;                                                                         \\
unassign('t'):unassign('l'):unassign('n'):unassign('u'):                    \\
close():                                                                    \\*
\rm

\subsubsection*{Listing 2}
\tt
  \#---------------------------------------------------------------------- \\
  \# Animation of the string instanton solutions. This program generates a \\
  \# .gif graphics file for each frame of the animation sequence.          \\
  \#                                                                       \\
  \# Bob Bacus c.1997                                                      \\
  \#------------------------- User input parameters ---------------------- \\
  p:=1:                 \# F= z\^p                                         \\
  q:=2:                 \# G=-z\^q                                         \\
  startframe:=0:        \# frame to start with                             \\
  endframe:=400:        \# frame to end with                               \\
  \#-----------------------------------------------------------------------\\
  space:=`         `:gc(300000):interface(plotdevice=gif):                     \\
  readlib(write):readlib(unassign):                                            \\
  for n from startframe by 1 to endframe do                                    \\
\hspace*{.2in}    d:=[]:                                                                     \\
\hspace*{.2in}    file:=cat(convert(p,string),`x`,convert(q,string),`-`,\\
\hspace*{1.0in}convert(n,string),`.dat`):\\
\hspace*{.2in}    indx:=parse(readline(file)):                                               \\
\hspace*{.2in}    if indx>=0 then tp:=cat(`p=`,convert(p,string),
  `  q=`,convert(q,string),`  Time Index
  :  `,substring(convert(indx,string),1..7),substring(space,1..7-length(substring(convert( \\
  indx,string),1..7)))) else tp:=cat(`p=`,convert(p,string),`  q=`,convert(q,string),`  \\
  Time Index:`,substring(convert(indx,string),1..8),substring(space,1..8-length(substring(\\
  convert(indx,string),1..8)))) fi:                                            \\
\hspace*{.2in}    points:=parse(readline(file)):                                             \\
\hspace*{.2in}    for i from 1 to q-1 do                                                     \\
\hspace*{.4in}      readline(file):                                                          \\
\hspace*{.4in}      j:=1:                                                                    \\
\hspace*{.4in}      for s from 1 to points/2 do                                              \\
\hspace*{.6in}        pt:=parse(readline(file)):                                             \\
\hspace*{.6in}        if s=1 then if linalg[iszero](map(Im,pt)) then                                  \\
  smin:=pt :b(j):=[pt] else b(j):=[]
  : smin:=[2.999,2.999,2.999] fi fi:                                           \\
\hspace*{.6in}        if s>1 then if linalg[iszero](map(Im,pt)) then 
b(j):=[op(b(j)),pt] else j:=j+1 :b  (j):=[]: fi fi:                                          \\
\hspace*{.4in}      od;                                                                      \\
\hspace*{.4in}      k:=1:                                                                    \\
\hspace*{.4in}      c(1):=[]:                                                                \\
\hspace*{.4in}      for s from points/2+1 to points do                                       \\
\hspace*{.6in}        pt:=parse(readline(file)):                                             \\
\hspace*{.6in}        if s<points then if linalg[iszero](map(Im,pt))  then c(k):=[op(c(k)),pt] \\
     else k:=k+1: c(k):=[]: fi fi:                                                         \\
\hspace*{.6in}        if s=points then if linalg[iszero](map(Im,pt)) then smax:=pt: c(k):=[op(c(k)),pt]\\
   else smax:=[2.999,2.999,2.999] fi fi:                                       \\
\hspace*{.4in}      od;                                                                      \\
\hspace*{.4in}      L1:=[]:                                                                  \\
\hspace*{.4in}      L2:=[]:                                                                  \\
\hspace*{.4in}      for h from 1 to j do                                                     \\
\hspace*{.6in}        if nops(b(h))>0 then L1:=[op(L1),b(h)] fi:                             \\
\hspace*{.4in}      od:                                                                      \\
\hspace*{.4in}      for h from 1 to k do                                                     \\
\hspace*{.6in}        if nops(c(h))>0 then L2:=[op(L2),c(h)] fi:                             \\
\hspace*{.4in}      od:                                                                      \\
\hspace*{.4in}      if nops(L1)>0 then L1:=op(1,L1): else L1:=[]: fi:                        \\
\hspace*{.4in}      if nops(L2)>0 then L2:=op(1,L2): else L1:=[]: fi:                        \\
\hspace*{.4in}
  d:=[op(d),PLOT3D(POINTS([2.999,2.999,2.999]),COLOR(RGB,0,0,0)),\\
PLOT3D(POINTS([smin,smax]),COLOR(RGB,0,0,0)),PLOT3D(CURVES(L1),COLOR(RGB,1,0,0)),PLOT3D(CURVES(L2),\\
COLOR(RG  B,0,0,1))]:                                                                  \\
\hspace*{.2in}    od:                                                                        \\
\hspace*{.2in}
  interface(plotoutput=cat(convert(p,string),`x`,convert(q,string),\\
`-`,convert(n,string  ),`.gif`)):                                                                  \\
\hspace*{.2in}    plots[display](d,view=[-3..3,-3..3,-3..3],projection=.1,orientation=[25,60],axes=boxe
  d,labels=['x','y','z'],scaling=constrained,titlefont=[COURIER,10],title=tp); \\
  test:=readline(file):                                                        \\
  unassign('l'):unassign('b'):unassign('points'):close():                      \\
  od:                                                                          \\
\rm

\section{Acknowledgements}  V.G.J. Rodgers thanks V.P. Nair for
discussion.

\section*{REFERENCES}
\begin{enumerate}
\item\label{bib:tHooft}G. t'Hooft, Nucl. Phys. B72 (1973) 461
\item\label{bib:Wilson}K. G. Wilson, Phys. Rev. D 10 (1974) 2445
\item\label{bib:Gervais}J.-L. Gervais and A. Neveu, Phys. Lett. B80
  (1979) 255
\item\label{bib:Makeenko}Yu M. Makeenko and A.A. Migdal,
  Phys. Lett. B88 (1979) 135
\item\label{bib:Belavin}A. A. Belavin, A. M. Polyakov, A. Schwartz, Y. Tyupkin,
Phys. Lett. 59B (1975) 85
\item\label{bib:Witten}E. Witten, Phys. Rev. Lett. 38 (1977) 121
\item\label{bib:Polyakov1}A. M. Polyakov, Nucl. Phys. B268 (1986) 406
\item\label{bib:Balachandran} A.P. Balachandran, F. Lizzi, G. Sparano, 
Nucl. Phys. B263 (1986) 608
\item\label{bib:Nair} P.O. Mazur and V.P. Nair, Nucl. Phys. B284
  (1986) 146
\item\label{bib:Durhuus} J. Ambjorn and B. Durhuus, Phys. Lett. B188
  (1987) 253
\item\label{bib:Parthasarathy} R.Parthasarathy and K.S.Viswanathan,
Lett.Math.Phys. 48 (1999) 243 
\item\label{bib:Wheater}J. F. Wheater, Phys. Lett. B208 (1988) 388
\item\label{bib:Robertson}G. Robertson, Phys. Lett. B 226 (1989) 244
\item\label{bib:Landolfi}  B.G. Konopelchenko and G. Landolfi,
 Phys.Lett. B459 (1999) 522
\item\label{bib:Pawelczyk} J.Pawelcyzk, Phy. Rev. Lett 74 (1995) 3924;
   Phys. Lett. B387 (1996) 287;  Nucl. Phys. B491, (1997) 515
\item\label{bib:Rosensweig}V.P. Nair and C. Rosenzweig,
  Phys. Lett. B135 (1984) 450; Phy. Rev. D31 (1985) 401
\item\label{bib:Christ}Allan S. Blaer, Norman H. Christ, and Ju-Fei
  Tang,
Phy. Rev. D25 (1982) 25
\item\label{bib:Callan} C.G. Callan, Jr.  Phys. Rev. D25 (1982) 2141
\end{enumerate}

\end{document}